\documentclass[11pt,a4paper]{article}
\usepackage{jheppub}
\usepackage{dsfont}
\usepackage{graphicx}
\usepackage{lipsum}
\usepackage{slashed}
\usepackage{hyperref}
\usepackage{mathtools}
\usepackage{bm}
\usepackage[all]{hypcap}
\usepackage{tikz}
\usepackage{tikz-feynman}
\usepackage{bbm}
\usepackage{pifont}
\usepackage{xcolor}
\usepackage{subcaption}
\usepackage[font=small,labelfont=bf]{caption}

\allowdisplaybreaks

\def\Z{\mathbb{Z}}

\def\g{\mathfrak{g}}

\def\su{\mathfrak{su}}

\usepackage{xspace}

\newcommand{\be}{\begin{equation}}
\newcommand{\ee}{\end{equation}}
\newcommand{\bea}{\begin{eqnarray}}
\newcommand{\eea}{\end{eqnarray}}
\newcommand{\beq}{\begin{equation}}
\newcommand{\eeq}{\end{equation}}

\newcommand{\cL}{{\cal L}}

\title{ In Search of an Invisible ${\bm Z^\prime}$ }

\author{Joe Davighi}
\emailAdd{joseph.davighi@cern.ch}

\affiliation{Theoretical Physics Department, CERN, 1211 Geneva 23, Switzerland}

\abstract{ 
We consider extending the Standard Model by an anomaly-free and possibly flavour non-universal $U(1)_X$ gauge symmetry, whose breaking gives a $Z^\prime$ boson that does not affect electroweak precision observables at tree-level or via 1-loop renormalisation group (RG) running. Provided it does not also couple to electrons, such a $Z^\prime$ boson would be largely invisible to an electroweak precision machine like FCC-ee or CEPC (up to small finite 1-loop matching contributions that we quantify). We show that, while this class of $Z^\prime$ models can also evade tests of quark flavour violation, the constraint of anomaly-cancellation implies that valence quarks, muons, and taus are all charged under $U(1)_X$, with the up quark charge being necessarily large. The conclusion holds even if one augments the SM by three right-handed neutrinos to try and absorb anomalies. This means such $Z^\prime$ bosons cannot simultaneously hide below the TeV scale from $pp \to \ell\ell$ Drell--Yan measurements at the LHC and, even if we entertain esoteric models in which the lepton charges are numerically very small, we cannot escape dijet searches at the LHC. For equitable quark and lepton charges, $pp\to\ell\ell$ already excludes such a $Z^\prime$ up to $M/g \gtrsim$ 10 TeV, with a reach of 20 TeV expected by the end of the High-Luminosity LHC. The dijet bounds currently sit around 5 TeV, while sensitivity up to 10 TeV could be achieved at HL-LHC. We thus find an excellent complementarity between FCC-ee and HL-LHC in covering all anomaly-free $Z^\prime$ bosons up to several TeV.
}

\begin{document}

\begin{flushright}
CERN-TH-2024-212 \end{flushright}

\maketitle

\section{Introduction}

In Ref.~\cite{Allwicher:2024sso} it was shown that very few single-particle extensions of the Standard Model (SM) can escape running into electroweak precision observables (EWPOs) at 1-loop level. The level of precision anticipated at FCC-ee, particularly on its $Z$-pole run, promises to probe all such states up to scales of several TeV, depending on the gauge and Lorentz representation of the particle.

The exceptions are:
(i) a scalar diquark field 
that is coupled only to top quarks; (ii) a vector $SU(2)_L$ triplet, or $W^\prime$, whose flavour-violating couplings to light quarks are precisely tuned against its flavour-conserving couplings; (iii) a vector $SU(3)$ octet coupled only to top quarks;
and (iv) a vector singlet, or $Z^\prime$, with couplings to fermions that satisfy certain linear equations.\footnote{Colourless scalar fields transforming in the ${\bf 4}$-plet of $SU(2)_L$, with hypercharge $1/2$ or $3/2$, do not contribute to the RGEs for SMEFT operators constrained by the $Z$-pole, but they {\em do} however generate finite 1-loop contributions to the $O_{HD}$ operator~\cite{Anisha:2021hgc,Durieux:2022hbu}, that shifts the $W$ mass. 
As shown in~\cite{Durieux:2022hbu}, these effects can be pushed to higher order (in loops or the $1/\Lambda$ expansion) by considering the custodially symmetric combination of these two scalar fields, which could originate from a scalar transforming in the ${\bf 16}$ of custodial $SO(4)$ symmetry. Such a model might leave its largest imprint on Higgs self-coupling measurements. \label{foot:4plet}
} 
Of these options, it is hard to imagine the $W^\prime$ with such tuned couplings to the SM fermions emerging from any reasonable UV model, and moreover it is already ruled out to $M/g \gtrsim 1000$ TeV by its flavour-violating impact on neutral meson mixing, making its accidental invisibility on the $Z$-pole a moot point.
The two coloured options, while tuned to escape the 1-loop RGEs, nonetheless exhibit a large-ish running into EWPOs at next-to-leading order that scales with the top Yukawa and the strong gauge coupling~\cite{Allwicher:2023aql} and, more importantly, they will already be probed up to several TeV by the LHC thanks to their direct couplings to gluons.
For the final case, namely the $Z^\prime$, there is a wide family of possible solutions, which can arise from a reasonable UV model in which the SM is extended by some gauged $U(1)_X$ symmetry.
Since the $Z^\prime$ is a gauge singlet, and its couplings can be flavour conserving, it is unclear {\em prima facie} whether any of these $Z^\prime$ models can simultaneously hide from other experiments, principally at the LHC. 

In this paper, we look to fill the remaining gap, by asking if any of these $Z^\prime$ models that are more-or-less invisible on the $Z$-pole\footnote{In fact, as for the scalar $SU(2)_L$ quadruplet discussed in footnote~\ref{foot:4plet}, the $Z^\prime$ models we construct (that give vanishing SMEFT RGE running into the $Z$-pole) do generate small finite 1-loop contributions to the $O_{Hf}$ operators, where $f\in\{e,u,d\}$, which we compute in \S \ref{sec:Zpole} to estimate the sensitivity of tera-$Z$ which, sure enough, does not reach a TeV. } 
can also escape constraints from current experiments and remain viable at a low-scale, let's say at $M/g \lesssim$ TeV, beyond FCC-ee. The answer is a resounding no. We find that the complementary constraints that already exclude such electroweakly-invisible $Z^\prime$ bosons come not from indirect flavour measurements, as anticipated in~\cite{Allwicher:2024sso}, but from direct searches at high-energy.

To show this, we will assume the $Z^\prime$ is the heavy gauge boson associated to a spontaneously broken $U(1)_X$ gauge symmetry.
Then the $U(1)_X$ charges of SM fermions must be chosen such that gauge anomalies cancel. This imposes a system of linear and non-linear constraints which must be solved over the rational numbers. 
We find that these constraints become extremely restrictive when combined with those conditions that result from requiring $Z$-pole invisibility. In particular, we show that:\footnote{
In a similar spirit Ref.~\cite{Greljo:2022dwn} found that anomaly cancellation when combined with constraints related to explaining the $(g-2)_\mu$ anomaly, such as it was in 2022, ends up being very predictive concerning the space of viable models.
}
\begin{itemize}
    \item Right-handed up and charm quarks are charged under $U(1)_X$
    \item At least two generations of right-handed down-type quarks are charged under $U(1)_X$
    \item At least two generations of right-handed leptons are charged under $U(1)_X$
    \item The up-quark charge is moreover {\em large}, necessarily satisfying $2|u| \geq \text{max}(|d_i|,|e_i|)$, meaning the $Z^\prime$ coupling to protons is unsuppressed.
\end{itemize}
All of these statements hold even if we consider augmenting the SM fermion content by arbitrarily many right-handed neutrinos charged under $U(1)_X$ to try and `soak up' gauge anomalies.

The result is that such a $Z^\prime$ model is, inevitably, strongly constrained already at the LHC, with current constraints already excluding $M/g$ up to 5 TeV. To reach this conclusion, we try to decouple as many further constraints as possible. Firstly, if right-handed electrons are charged, then there are also strong constraints coming from LEP-II $e^+e^-\to l^+l^-$ cross-section measurements. 
As highlighted in recent studies~\cite{Ge:2024pfn,Greljo:2024ytg}, the power of the electroweak programme at FCC-ee is not limited to extreme precision on the $Z$-pole:
we infer that the $Z^\prime$ models we consider, if coupled to $e_R$, would be probed up to 30 TeV  at FCC-ee by measurements {\em off} the $Z$-peak. By assigning the two non-zero lepton charges to muon and tau, and demanding the electron charge be zero, these constraints from $e^+e^-$ colliders such as FCC-ee disappear. In this case, we provide an analytic parametrization of all the viable $U(1)_X$ charge solutions, which ends up equivalent to solving for Pythagorean triples.

Continuing with our search for an invisible $Z^\prime$, there remain constraints coming from LHC $pp \to \ell\ell$ Drell--Yan measurements, which are very strong especially for muons in the final state and given that up and down quarks must have unsuppressed couplings to the $Z^\prime$ in order to not run into the EWPOs.
We find current LHC data from $pp \to \ell\ell$ at high-$p_T$ requires $M/g \gtrsim 10$ TeV, which should improve to reach roughly 20 TeV after the High-Luminosity LHC (HL-LHC). Finally, one can try to weaken also these $pp \to \ell\ell$ constraints by finding viable charge assignments in which the lepton charges are numerically much smaller than the quark charges. But even in this case, we cannot help running into di-jet searches at the LHC, which give very strong constraints within the kinematical reach of the searches which extends up to 5 TeV. 
We give a crude estimate of the bounds from di-jet resonance searches, which apply only for narrow-width $Z^\prime$s, using Ref.~\cite{Bordone:2021cca}, and also estimate bounds from non-resonant searches performed using angular distributions of dijets~\cite{CMS:2018ucw}.

The outline of the paper is as follows. In \S \ref{sec:neutrality} we recap the conditions for a $Z^\prime$ to not run into the EWPOs from~\cite{Allwicher:2024sso}. 
In \S \ref{sec:ACC} we review the conditions on a $Z^\prime$ to arise from an anomaly-free $U(1)_X$ extension of the SM, and highlight an interplay between these conditions and those from not running into the $Z$-pole. In \S \ref{sec:NT} we systematically explore the space of $Z^\prime$ models satisfying these criteria, deriving the various statements sketched in the Introduction. Then in \S \ref{sec:pheno} we sketch the main phenomenological bounds on the models, coming from $4$-lepton, semi-leptonic, and $4$-quark probes. We compare various benchmark models, and those that {\em do} run into the EWPOs, in Fig.~\ref{fig:BAR}, before concluding in \S \ref{sec:conc}.

\section{Electroweak neutrality} \label{sec:neutrality}

We assume the $Z^\prime$ comes from a gauged $U(1)_X$ extension of the SM, and we label the charges of the various SM fields by $\{q_i, l_i, u_i, d_i, e_i, H\}$ in what we hope is an obvious notation. For simplicity, we suppose the $U(1)_X$ symmetry is spontaneously broken by a SM singlet scalar field $\phi$, with unit charge under $U(1)_X$, that gets a vev $ \langle \phi \rangle = v_X/\sqrt{2}$, resulting in the $Z^\prime$ getting a mass 
$M_X = g_X v_X$.

\subsection{Constraints from 1-loop SMEFT running into the $Z$-pole}

We begin by recapping the linear constraints on the $U(1)_X$ charges that should be satisfied to not run into EWPOs via the one-loop level RGEs, which was studied in~\cite{Allwicher:2024sso}. The SMEFT operators that effect the EWPOs at tree-level are, in the Warsaw basis,
\begin{equation} \label{eq:EWPO_SMEFT}
    \{[O_{H\ell}^{(1,3)}]_{ii}, [O_{Hq}^{(1,3)}]_{ii}, [O_{He}]_{ii}, [O_{Hu}]_{ii} \,(i=1,2\text{~only}), [O_{Hd}]_{ii}, O_{HD}, O_{HWB}, [O_{\ell\ell}]_{1221} \}\, .
\end{equation}
For the $Z^\prime$ gauge boson to not shift EWPOs already at tree-level requires the Higgs is neutral, 
        $H=0$,
and that the $Z^\prime$ has no flavour off-diagonal coupling to left-handed muon-electron bilinears. This can be achieved either via a universal coupling to both ($l_1=l_2$) or by alignment between the electron and muon mass and flavour eigenstates. We see shortly that cancelling also the one-loop running into \eqref{eq:EWPO_SMEFT} implies the former scenario.

At one-loop and at leading order in the SMEFT Wilson coefficients, many additional SMEFT operators run into the set (\ref{eq:EWPO_SMEFT}) via the SMEFT RGEs. Neglecting all Yukawa couplings other than $y_t$, and keeping the running with all three SM gauge couplings $g_{1,2,3}$, it is straightforward to identify the conditions for the $Z^\prime$ to not run into the EWPOs, in other words to be `electroweakly neutral' at this order. Recapping from~\cite{Allwicher:2024sso}:
\begin{itemize}
    \item 
    Running with the top Yukawa coupling $y_t$ vanishes iff 
        $q_3=u_3$.
    This is the condition for $U(1)_X$ permitting a renormalisable top Yukawa coupling, which we want to enforce anyway;
    \item 
    Running with the $SU(2)_L$ gauge coupling $g_2$ vanishes iff all left-handed fermion fields are neutral under $U(1)_X$, {\em i.e.} 
        $l_i = q_i = 0 \, \forall i\in\{1,2,3\}$.
    Note this implies $[C_{\ell\ell}]_{1221}$ is not generated at tree-level.
    Together with the previous condition, this also implies
            $u_3=0$,
    so the top quark is neutral under $U(1)_X$;
    \item 
    Lastly, running with the $U(1)_Y$ gauge coupling $g_1$ vanishes iff the further linear condition on the remaining non-zero charges is satisfied:
    \begin{equation} \label{eq:g1_run}
        \sum_i \left(-e_i +2u_i-d_i \right) = 0\, .
    \end{equation}
\end{itemize}
We discuss the implications of this final condition when we turn to anomaly cancellation.

At this point, we already wish to emphasize that the condition from $g_2$ running means these $Z^\prime$ models are intrinsically chiral, coupling only to right-handed fermion fields. While anomaly-cancellation is still possible (see \S \ref{sec:ACC}), this does however immediately imply that full Yukawa coupling matrices are not permitted at the renormalisable level. Rather, only certain elements of the Yukawa matrices are permitted at dimension-4; encouragingly, the top Yukawa coupling $\overline{q_3} H^c u_3$ is always permitted thanks to the top quark and Higgs being neutral, which was implied by requiring no 1-loop running into the $Z$-pole with $y_t$ or $g_3$. But otherwise, some additional dynamics is needed to populate the full Yukawa matrices, and one can push this to a higher scale without destabilising the lowest layer of new physics that we assume is associated to the $Z^\prime$.\footnote{Viewed from a different perspective, the chiral and flavour non-universal nature opens up the possibility that $U(1)_X$ can explain aspects of the flavour puzzle: some Yukawa couplings must originate from higher-dimensional operators which are therefore suppressed with respect to the renormalisable top quark Yukawa coupling. But our focus here is on the phenomenology of the $Z^\prime$ and not on solving the flavour puzzle. }

We also remark that, while we have ensured the cancellation of SMEFT RGE contributions to the EWPOs, this does not preclude the existence of finite (non logarithmically-enhanced) 1-loop matching contributions to the EWPOs that would still lead to some weak sensitivity of the tera-$Z$ run to these $Z^\prime$ models. We turn to these finite matching contributions in \S \ref{sec:Zpole}. Indeed, we will see that, for all the `electroweakly neutral' $Z^\prime$ models we arrive at, the EWPO constraints coming from these finite one-loop effects are weaker than other complementary constraints. 

\subsection{Electroweak neutrality beyond the $Z$-pole} \label{sec:neutrality-offpeak}

As mentioned in the Introduction, the electroweak program of FCC-ee extends beyond the virtues of its tera-$Z$ run, as emphasized for instance in the studies~\cite{Ge:2024pfn,Greljo:2024ytg} (see also~\cite{deBlas:2022ofj,Celada:2024mcf,Maltoni:2024csn}). Of particular relevance to our $Z^\prime$ models, FCC-ee measurements above the $Z$-peak will put strong constraints on 4-fermion operators involving electrons. 

The constraints on such operators coming from LEP-II, in particular from  $e^+e^- \to \ell^+\ell^-$ cross-section and forward-backward asymmetry measurements, are already strong and probe a model in which the electron is charged up to scales of several TeV (as we show for our class of $Z^\prime$ models in \S \ref{sec:4l}).
At FCC-ee, the bounds on dimension-6 4-lepton operators containing electrons will jump to $20\div 30$ TeV~\cite{Greljo:2024ytg}.
Therefore, in our hunt for a light $Z^\prime$ model that could escape detection at FCC-ee, it had better not couple to electrons either, and so we should restrict to solutions in which $e_1=0$ also. We do so in \S \ref{sec:one-lepton}.

\subsection{Flavour Neutrality}

The condition that all left-handed quarks and leptons are neutral under $U(1)_X$ means that such models can evade most constraints coming from flavour violation probes. This is because, while for left-handed quarks we know there is non-zero misalignment between the flavour and gauge eigenstates in order to reproduce the CKM matrix, the right-handed quarks can be aligned. If only right-handed fields are charged under the $Z^\prime$, this means that models can be consistently constructed in which the $Z^\prime$ does not induce any flavour violation. More realistically, the quark flavour violation will be suppressed by the size of mixing angles amongst the right-handed quark fields, which must be sufficiently small.\footnote{Small right-handed mixing angles are a face-value expectation from the gauged $U(1)_X$ symmetries that we end up considering. This is because right-handed fields acquire non-universal charges under $U(1)_X$ and so in general we expect different EFT suppression factors appearing in each column of the Yukawa matrices, resulting in hierarchically suppressed right-handed mixing. See \S \ref{sec:flavour-structure}.
} 
The same goes for leptons. So, this class of $Z^\prime$ models that are invisible to electroweak precision have a consistent limit in which they are invisible also to tests of flavour violation. This means that flavour-conserving constraints on 4-fermion operators will dominate the phenomenology, as we will see in \S \ref{sec:pheno}.

\section{Anomaly cancellation} \label{sec:ACC}

The $U(1)_X$ extension of the SM should also be anomaly-free.
In a $U(1)_X$ extension of the SM there are six possible triangle anomalies involving $U(1)_X$ gauge bosons and SM gauge bosons, including the mixed anomaly between $U(1)_X$ and gravity. Requiring these perturbative anomalies cancel\footnote{
In a $U(1)_X$ extension of the SM it is sufficient to consider only the cancellation of perturbative gauge anomalies. As long as we do not add an odd number of chiral fermions with $SU(2)_L$ isospin $j \in 2\Z+1/2$, there are no non-perturbative gauge anomalies to further consider~\cite{Davighi:2019rcd,Wan:2019gqr}.
} results in six homogeneous polynomial equations that must be solved over the rationals: four are linear, one is quadratic, and one is cubic in the charges. One of the linear conditions, namely the mixed anomaly with $SU(2)_L$, is automatically satisfied because we already learnt that all $SU(2)_L$ doublets must be $U(1)_X$ neutral to avoid the running into EWPOs with $g_2$.

Before we get started solving these equations, the reader might object that, since the $U(1)_X$ gauge symmetry is necessarily spontaneously broken, it is possible for gauge anomalies to be cancelled in the low-energy EFT by Wess--Zumino--Witten (WZW) terms~\cite{Wess:1971yu,Witten:1983tw} \`a la Preskill~\cite{Preskill:1990fr}, that we suppose arise from integrating out some heavy chiral fermions that restore anomaly cancellation in the underlying renormalisable theory.
But it is reasonable to discard this class of EFTs in our pursuit of a $Z^\prime$ that is invisible on the $Z$-pole, as follows. 
Firstly, if the heavy fermions that cancel anomalies are SM singlets, they can at most cancel the pure 't Hooft anomaly associated with gauging $U(1)_X$ and the mixed anomaly between $U(1)_X$ and gravity. Such states are simply heavy right-handed neutrinos, and we discuss this extension of the SM field content explicitly in \S \ref{sec:neutrino}, finding that our main conclusions go through unchanged.
Secondly, if we entertain heavy chiral fermions that cancel mixed anomalies between $U(1)_X$ and the SM, then they must be charged (although not necessarily chirally) under the SM gauge group. If they are electroweakly-charged, as in {\em e.g.} Ref.~\cite{Davighi:2021oel}, the extra chiral fermions couple directly to the electroweak gauge bosons and so give 1-loop contributions to the EWPOs anyway, so we should strike out this option in our pursuit of an invisible $Z^\prime$. If they are coloured, they radically alter the LHC phenomenology of such a model, with the heavy quarks likely providing the leading signature to target rather than the $Z^\prime$.\footnote{Note that, precisely because the heavy quark is posited to cancel a mixed anomaly between colour and $U(1)_X$, we cannot make it arbitrarily heavier than the $Z^\prime$, without simultaneously making the $Z^\prime$ arbitrarily weakly coupled to the SM fermions.} 

We therefore stick to the simpler scenario in which the $U(1)_X$ charges are chosen so that all gauge anomalies cancel given the SM chiral fermion contributions only, with the exception of possible right-handed neutrino contributions that we treat in \S \ref{sec:neutrino}.

\subsection{Linear anomaly conditions}

Given the mixed anomaly with $SU(2)_L$ is automatically satisfied, there are three linear anomaly cancellation conditions (ACCs) to solve, plus the two non-linear ones. Allowing a right-handed electron charge for now, recall there are eight rational charges left to fix, namely $\{u_1, u_2, d_i, e_i\}$ where $i\in\{1,2,3\}$. To proceed, it is convenient to define the sums of charges:
\begin{equation}
    D = \sum_i d_i, \qquad E = \sum_i e_i\, ,
\end{equation}
since it is these sums that appear in all the linear anomaly conditions, given their invariance under permuting family indices. Note the `$g_1$ running' condition (\ref{eq:g1_run}) is then
    $2(u_1 + u_2) = E + D$.
The mixed anomalies with $SU(3)_c$ and $U(1)_Y$ (the one with a single $X$ leg) vanish iff
    $u_1+u_2 = -D$ and
    $8(u_1+u_2) = -2D-6E$ 
respectively, which implies $E=D=-(u_1+u_2)$.
The $g_1$ running condition is then only satisfied if
\begin{equation}
    u_1+u_2 = 0 \quad \Rightarrow \quad E=D=0\, .
\end{equation}
All the linear constraints are now satisfied. Note that the mixed anomaly with gravity automatically vanishes without needing to enforce it as a constraint, because the other conditions already imply that the sums of charges for every fermion type equal zero. 

One can also flip the logic, and consider first imposing all the linear ACCs including the mixed anomaly with gravity. It is known that this forces the `sums of charges' $D$, $E$ {\em etc} to be a linear combination of hypercharge and $B-L$, and so knowing also that all the left-handed fields are neutral immediately imply all sums of charges are zero. Then the 1-loop running with $g_1$ into $Z$-pole observables cancels exactly upon summing over flavours of each type of fermion. For instance, the contributions of 4-lepton SMEFT operators $O_{ee}^{iijj}$ that arise from integrating out the $Z^\prime$ into the Higgs-bilepton current are
\begin{equation} \label{eq:feynman}
    \dot C_{He}^{ii} \sim \sum_{j=1}^3\,\,
        \begin{tikzpicture}[baseline=-3.5ex]
        \begin{feynman}
        \vertex (a) {$e_i$};
        \vertex [right=0.6in of a] (b);
        \vertex [right=0.6in of b] (c);
        \vertex [below=0.4in of a] (e)  {${e}_i$};
        \vertex [below=0.2in of b] (f);
        \node at (f) [square dot,fill,inner sep=1.5pt]{};
        \vertex [below=0.2in of c] (g);
        \node at (g) [circle,fill,inner sep=1.5pt]{};
        \vertex [right=0.5in of g] (h) {$Z_\mu$};
        \diagram{
        (a) -- [](f), 
        (f) -- [](e), 
        (g) -- [photon](h),
        (f) -- [half left, edge label'=$e_j$] (g) -- [half left] (f)
        };
        \end{feynman}
        \end{tikzpicture}
        \,\,
    \propto \,\, g_1^2 \sum_{j=1}^3 e_j = 0 \quad \text{by ACCs} ,
\end{equation}
where a square dot denotes a dimension-6 operator obtained from integrating out the $Z^\prime$ at tree-level, here a 4-fermion operator with $(\bar{R}R)(\bar{R}R)$ chiral structure, while a circular dot denotes a SM coupling, in this case a gauge coupling.
Separate cancellations occur when summing over down-type and up-type quarks in the loop.
The same goes for the $O_{Hu}$ and $O_{Hd}$ Higgs-fermion bilinears at 1-loop.
Thus the modifications of $Z$ couplings to right-handed fermions, sometimes denoted $\delta g^{Zf}_{R\, ii}$ for $f\in \{u,d,e\}$, vanish at this order in the SMEFT and loop expansion.

This diagrammatic argument also makes it clear that the cancellations to the RGEs are exact only because we are neglecting the masses of the leptons and quarks running inside the loop. This assumption corresponds to dropping the dependence of the SMEFT RGEs on all Yukawa couplings other than $y_t$ in Ref.~\cite{Allwicher:2024sso}. We infer that the leading RGE-induced corrections to $Z$ couplings will be highly suppressed by the ratio $m_{b,c,\tau}^2/M_X^2$, noting that there are no contributions with tops running in the loop because $u_3=q_3=0$. 
Comparable contributions would likely come from including dimension-8 operators for a TeV scale $Z^\prime$, given ${m_b m_X}/{m_Z^2} \sim \left(m_X/{2\text{~TeV}}\right)^2$.


\subsection{Corollary: zero 1-loop RGE with $y_t$ and $g_i$} \label{sec:RG}

This non-renormalisation at 1-loop of the $Z$ boson current is an example of a more general feature of the class of $Z^\prime$ models we have arrived at.
The 1-loop RGEs, keeping only effects in $y_t$ and the gauge couplings $g_i$, in fact vanish completely for {\em all} dimension-6 SMEFT operators, not just the subset \eqref{eq:EWPO_SMEFT} that enter the EWPOs at tree-level. It is again straightforward to see this by considering 1-loop diagrams -- or by inspecting the complete formulae for 1-loop RGEs derived in Refs.~\cite{Jenkins:2013zja,Jenkins:2013wua,Alonso:2013hga}. 

\paragraph{Yukawa dependence.}

Firstly, let us consider running of any SMEFT operators with $y_t$ in these models. Because the charges of $u_3$, $q_3$, and $H$ are all required to be zero by our electroweak neutrality conditions, we do not generate any dimension-6 SMEFT operators at tree-level that can connect to a top Yukawa insertion, meaning all the 1-loop diagrams involving one SMEFT coefficient and insertions of the top Yukawa vanish. For example, 
\begin{equation} \label{eq:yt-loop}
    \begin{tikzpicture}[baseline=-4.0ex]
    \begin{feynman}
        \vertex (a);
        \node at (a) [square dot,fill,inner sep=1.2pt]{};
        \vertex [above=0.3in of a] (aa);
        \vertex [right=0.15in of aa] (b) {${f}_i$};
        \vertex [left=0.15in of aa] (c) {${f}_i$};
        \vertex [below=0.3in of a] (d);
        \vertex [right=0.3in of d] (e);
        \node at (e) [circle,fill,inner sep=1.2pt,label=\footnotesize{$\,\,\,\,\,y_t$}]{};
        \vertex [left=0.3in of d] (f);
        \node at (f) [circle,fill,inner sep=1.2pt,label=\footnotesize{$y_t\,\,\,\,\,$}]{};
        \vertex [right=0.3in of e] (g) {${q}_3$};
        \vertex [left=0.3in of f] (h) {${q}_3$};
        \diagram* {
        (a) -- (b), 
        (a) -- (c),
        (a) -- [quarter left, edge label=\footnotesize{${u}_3$}] (e), 
        (a) -- [quarter right, edge label'=\footnotesize{$u_3$}] (f),
        (e) -- [scalar, half left, edge label=$H$] (f), 
        (e) -- (g),
        (f) -- (h)
        };
    \end{feynman}
    \end{tikzpicture}
\quad = \quad 
    \begin{tikzpicture}[baseline=-4.0ex]
    \begin{feynman}
        \vertex (a);
        \node at (a) [square dot,fill,inner sep=1.2pt]{};
        \vertex [above=0.3in of a] (aa);
        \vertex [right=0.15in of aa] (b) {${f}_i$};
        \vertex [left=0.15in of aa] (c) {${f}_i$};
        \vertex [below=0.3in of a] (d);
        \vertex [right=0.3in of d] (e);
        \node at (e) [circle,fill,inner sep=1.2pt,label=\footnotesize{$\,\,\,\,\,y_t$}]{};
        \vertex [left=0.3in of d] (f);
        \node at (f) [circle,fill,inner sep=1.2pt,label=\footnotesize{$y_t\,\,\,\,\,$}]{};
        \vertex [right=0.3in of e] (g) {$H$};
        \vertex [left=0.3in of f] (h) {$H$};
        \diagram* {
        (a) -- (b), 
        (a) -- (c),
        (a) -- [quarter left, edge label=\footnotesize{${u}_3$}] (e), 
        (a) -- [quarter right, edge label'=\footnotesize{$u_3$}] (f),
        (e) -- [half left, edge label=$q_3$] (f), 
        (e) -- [scalar] (g),
        (f) -- [scalar] (h)
        };
    \end{feynman}
    \end{tikzpicture}
\quad  =0
\end{equation}
because the 4-fermion operator vanishes at tree-level, and likewise for any dimension-6 SMEFT operator involving Higgses.
Thus, if we drop all Yukawa couplings other than $y_t$, there is no 1-loop running with Yukawas for these models from the matching scale (at which the $Z^\prime$ is integrated out) down to the electroweak scale.

The leading RG running with Yukawa couplings will, for generic models in our class, be due to the bottom quark Yukawa $y_b \sim 0.01$ (using the value of $y_b$ at a scale $\mu=1$ TeV~\cite{Xing:2007fb}, assuming SM RGE up to that scale). 
This generates $y_b^2$-suppressed running into 4-fermion operators with $(\bar{L}L)(\bar{R}R)$ chirality structure, from the diagrams
\begin{equation}
    \begin{tikzpicture}[baseline=-4.0ex]
    \begin{feynman}
        \vertex (a);
        \node at (a) [square dot,fill,inner sep=1.2pt]{};
        \vertex [above=0.3in of a] (aa);
        \vertex [right=0.15in of aa] (b) {${f}_i$};
        \vertex [left=0.15in of aa] (c) {${f}_i$};
        \vertex [below=0.3in of a] (d);
        \vertex [right=0.3in of d] (e);
        \node at (e) [circle,fill,inner sep=1.2pt,label=\footnotesize{$\,\,\,\,\,y_b$}]{};
        \vertex [left=0.3in of d] (f);
        \node at (f) [circle,fill,inner sep=1.2pt,label=\footnotesize{$y_b\,\,\,\,\,$}]{};
        \vertex [right=0.3in of e] (g) {${q}_3$};
        \vertex [left=0.3in of f] (h) {${q}_3$};
        \diagram* {
        (a) -- (b), 
        (a) -- (c),
        (a) -- [quarter left, edge label=\footnotesize{${d}_3$}] (e), 
        (a) -- [quarter right, edge label'=\footnotesize{$d_3$}] (f),
        (e) -- [scalar, half left, edge label=$H$] (f), 
        (e) -- (g),
        (f) -- (h)
        };
    \end{feynman}
    \end{tikzpicture}
\quad \implies \dot{C}_{qf}^{33ii} \sim y_b^2 \frac{g_X^2}{M_X^2}  X_{d_3} X_{f_i} \qquad f\in \{u,d,e\}\, , 
\end{equation}
This would give, for instance, a very small contribution to $e^+e^- \to b\bar{b}$ but only if $e_1$ were charged.
The more important contribution, phenomenologically, would likely be $y_b$-suppressed running into $Z$-pole observables, due to
\begin{equation}
    \begin{tikzpicture}[baseline=-4.0ex]
    \begin{feynman}
        \vertex (a);
        \node at (a) [square dot,fill,inner sep=1.2pt]{};
        \vertex [above=0.3in of a] (aa);
        \vertex [right=0.15in of aa] (b) {${f}_i$};
        \vertex [left=0.15in of aa] (c) {${f}_i$};
        \vertex [below=0.3in of a] (d);
        \vertex [right=0.3in of d] (e);
        \node at (e) [circle,fill,inner sep=1.2pt,label=\footnotesize{$\,\,\,\,\,y_b$}]{};
        \vertex [left=0.3in of d] (f);
        \node at (f) [circle,fill,inner sep=1.2pt,label =\footnotesize{$y_b\,\,\,\,\,$}]{};
        \vertex [right=0.3in of e] (g) {$H$};
        \vertex [left=0.3in of f] (h) {$H$};
        \diagram* {
        (a) -- (b), 
        (a) -- (c),
        (a) -- [quarter left, edge label=\footnotesize{${d}_3$}] (e), 
        (a) -- [quarter right, edge label'=\footnotesize{$d_3$}] (f),
        (e) -- [half left, edge label=$q_3$] (f), 
        (e) -- [scalar] (g),
        (f) -- [scalar] (h)
        };
    \end{feynman}
    \end{tikzpicture}
\quad \implies \dot{C}_{Hf}^{ii} \sim y_b^2 \frac{g_X^2}{M_X^2}  X_{d_3} X_{f_i} \qquad f\in \{u,d,e\}\, , 
\end{equation}
leading to very small RG-induced modifications to the $Z$-boson couplings, including $Z\to b_R b_R$, $Z\to \mu_R\mu_R$, and $Z\to \tau_R\tau_R$ measured at the $Z$-pole.
Given our $Z^\prime$ couples at tree-level only to right-handed fermions, there are no other diagrams to consider for running into any operators due to Yukawa dependence.

Small as these contributions already are,
we shall see in \S \ref{sec:one-lepton} that the class of solutions that decouple from FCC-ee both on {\em and off} the $Z$-pole necessarily has one $d_i$ vanishing, and it is attractive to take this to be $d_3=0$ due to the flavour structure implied by $U(1)_X$ (\S \ref{sec:flavour-structure}). In those models, the leading Yukawa running is now quadratic in $y_c\sim 3\times 10^{-3}$, giving completely negligible effects. 

\paragraph{Gauge coupling dependence.}

The running with the gauge couplings can be captured via diagrams like that we already discussed in \eqref{eq:feynman}, which gives the modification to the gauge boson currents from which one can obtain the running into 4-fermion operators and Higgs-bifermion operators. Again these contributions cancel exactly upon setting all Yukawas other than $y_t$ to zero, although the cancellation is qualitatively different in the two cases. For the Yukawa running, we saw the largest contributions (the $y_t$ running) were simply set to zero by choosing $U(1)_X$ charge assignments that decouple from the Higgs and top; for the gauge running, on the other hand, the $U(1)_X$ charge assignment leads to a precise cancellation between non-vanishing one-loop diagrams in the SMEFT, like that in \eqref{eq:feynman}. 
This cancellation follows from flavour universality of the SM gauge interactions, combined with the strict flavour {\em non}-universality of the $U(1)_X$ force, in particular the `tracelessness' conditions $D=E=U=0$ that followed from anomaly cancellation plus our $Z$-pole running criteria. 
This general argument shows that it is not only the $g_i$-dependent running into the Higgs-bifermions operators constrained on the $Z$-pole that vanishes at one-loop, but also the $g_i$-dependent running into 4-fermion operators. This applies even for those 4-fermion operators that are generated at tree-level, {\em viz.}
\begin{equation} \label{eq:4f_renorm}
    \dot C_{ff^\prime}^{iijj} \sim 
    \sum_{f^{\prime\prime} \in \{u,d,e\}} \sum_{k=1}^3\,\,
        \begin{tikzpicture}[baseline=-3.5ex]
        \begin{feynman}
        \vertex (a) {$f_i$};
        \vertex [right=0.6in of a] (b);
        \vertex [right=0.6in of b] (c);
        \vertex [below=0.4in of a] (e)  {${f}_i$};
        \vertex [below=0.2in of b] (f);
        \node at (f) [square dot,fill,inner sep=1.5pt]{};
        \vertex [below=0.2in of c] (g);
        \node at (g) [circle,fill,inner sep=1.5pt]{};
        \vertex [right=0.5in of g] (h);
        \vertex [right=0.3in of h] (hh);
        \vertex [above=0.2in of hh] (i) {$f^{\prime}_j$};
        \vertex [below=0.2in of hh] (j) {$f^{\prime}_j$};
        \diagram{
        (a) -- [](f), 
        (f) -- [](e), 
        (g) -- [photon](h),
        (f) -- [half left, edge label=$f^{\prime\prime}_k$] (g) -- [half left] (f),
        (h) -- (i),
        (h) -- (j)
        };
        \end{feynman}
        \end{tikzpicture}
        \,\,
    = 0 \, . 
\end{equation}
As above, the cancellations are only perfect because we neglect the mass of all the SM fermions running in the loop, and so there will be small corrections to the $g_i$ running that scale with the light Yukawas.

We have established that, in the limit of vanishing light fermion Yukawas, there is no 1-loop RGE running of these $Z^\prime$ models from the matching scale down to the relevant scale of measurement. There are, however, finite parts of 1-loop diagrams that can be computed in the full UV theory which contribute to EWPOs, which are not captured by RGE running within the SMEFT. 
We derive these contributions, and the resulting constraints from EWPOs, in \S \ref{sec:Zpole}.

\subsection{Non-linear anomaly conditions}

After this brief digression, we continue our discussion of anomaly cancellation. There remain two non-linear conditions on the charges that should be satisfied for gauge anomalies to cancel. 
The non-linearity here means that the conclusions we can deduce from enforcing these constraints, which requires some elementary number theory, are somewhat less obvious, but no less powerful.

First, the mixed anomaly with hypercharge corresponding to a triangle diagram with two $X$ legs, that is quadratic in charges, vanishes iff 
\begin{equation} \label{eq:quad}
    4 u_1^2 = \sum_i d_i^2 + e_i^2\, ,
\end{equation}
where we used $u_2=-u_1$.
Secondly, the cubic anomaly, corresponding to a triangle diagram with three $X$ legs, vanishes iff 
\begin{equation} \label{eq:cubic}
     0 = \sum_i 3d_i^3+e_i^3\, ,
\end{equation}
where we have used the fact that the linear constraints already imply $u_1^3+u_2^3=u_1^3+(-u_1)^3=0$. In summary, we need to solve the non-linear equations (\ref{eq:quad}) and (\ref{eq:cubic}) subject to the linear constraints $\sum_i d_i = \sum_i e_i=0$.

\section{Classifying $Z^\prime$ models with electroweak neutrality} \label{sec:NT}

\subsection{Valence quarks are charged}

We can deduce some immediate consequences of the non-linear ACCs. The right-hand-side of eq. (\ref{eq:quad}) is a sum of squares which implies the left-hand-side is non-negative. Hence, a non-sterile\footnote{We define a `sterile solution' to be one in which all $U(1)_X$ charges of SM fields are zero, in which case the $Z^\prime$ boson would decouple modulo a possible kinetic mixing with the hypercharge gauge boson. If the kinetic mixing parameter $\epsilon X_{\mu\nu}B^{\mu\nu}$ is turned on, the $Z^\prime$ mimics a heavy copy of the SM hypercharge gauge boson, picking up couplings to the Higgs and all SM fermions after diagonalising the gauge field kinetic terms. The effective scale in that case would be $\Lambda = M/(\epsilon g_X)$, with the combination $\epsilon g_X$ being the effective coupling, and this $Z^\prime$ would be probed up to scales of order 80 TeV~\cite{Allwicher:2024sso} by the FCC-ee $Z$-pole run due to the direct couplings to all SM fields. Of course, such a $Z^\prime$ could also be very light ($M\ll M_Z$) and weakly-coupled, in which case it would serve as a dark photon -- a phenomenologically different beast to the heavy, strongly coupled extensions of the SM that we are concerned with here. } 
solution requires
\begin{equation} \label{eq:u1}
    u_1 \neq 0, \qquad u_2=-u_1\, .    
\end{equation}
Thus, both the right-handed up and charm quarks are unavoidably charged under $U(1)_X$. The coupling to up quarks in particular means that production of the $Z^\prime$ boson in $pp$ collisions has no flavour suppression, and so there will be strong constraints from the LHC.

Also notice that (\ref{eq:quad}) does not only imply that the up quark has non-zero charge, but also that its charge is {\em large}, because:
\begin{equation}
    2|u| \geq \text{max}(|d_i|,|e_i|)\, ,
\end{equation}
{\em i.e.} the up charge (multiplied by two) is at least as big as the largest lepton or down-type quark charge. It follows immediately that the $Z^\prime$ has large coupling to protons and so is abundantly produced in a proton-proton collider.

Knowing this, the next most important thing to know phenomenologically is whether the $Z^\prime$ also couples to charged leptons, and how many flavours thereof. If it does, the $Z^\prime$ will contribute to Drell--Yan production of di-lepton pairs at the LHC (\S \ref{sec:2q2l}). And even if it does not, or if the lepton charges can be numerically suppressed, condition (\ref{eq:u1}) implies the $Z^\prime$ will contribute to di-jet production at the LHC, which gives strong constraints for a $Z^\prime$ coupled to up quarks~\cite{Bordone:2021cca} as is the case here (\S \ref{sec:4q}). 

We also emphasize that the up-type quarks necessarily have all different charges under $U(1)_X$, {\em viz.} $(u_1, u_2=-u_1, u_3=0)$. This means there are no $Z^\prime$ models with $U(2)_u$ quark flavour symmetry, which would require $u_1=u_2$, that satisfy our criteria for electroweak neutrality. But, as discussed in \S \ref{sec:neutrality}, this does not necessarily mean there are strong bounds from {\em e.g.} kaon mixing or other precision flavour probes, because the strong $U(2)$-breaking is only present for right-handed quarks and there the mixing angles can be consistently tuned to be arbitrarily small.

\subsection{At least two leptons are charged} \label{sec:at-least-two-e}

It is quick to show that at least one right-handed lepton is necessarily charged, or else the $U(1)_X$ is sterile {\em i.e.} all charges are zero. To see this, let us try setting all three RH lepton charges $e_i$ to zero. The cubic anomaly constraint becomes 
\begin{equation}
    0 = d_1^3+d_2^3+d_3^3\, .
\end{equation}
Rather famously, this equation has no rational solutions in which all $d_i$ are non-zero, but rather the only solutions satisfy $d_1 d_2 d_3=0$, of the form $d_1=-d_2$, $d_3=0$ (up to permuting flavour indices). Substituting this into the quadratic, together with $e_i=0$, we get
\begin{equation}
    2u_1^2=d_1^2\, ,
\end{equation}
which has no rational solutions. Thus, by contradiction, we establish that at least one $e_i$ charge must be non-zero for there to exist rational solutions. But then the linear condition $E=\sum_i e_i=0$ can only be satisfied if at least {\em two} $e_i$ be non-zero.
Thus, in addition to coupling to up and charm quarks, an anomaly-free $Z^\prime$ that does not run into the $Z$-pole observables necessarily couples at tree-level to (at least) two flavours of leptons. This combination of couplings means there are strong constraints coming from LHC Drell--Yan data at high-$p_T$ (\S \ref{sec:2q2l}), including at least one light flavour lepton channel. 

\paragraph{At least two down-type quarks are charged.} 
The same mathematical argument used in \S \ref{sec:at-least-two-e} applies to the down quarks: that is, (\ref{eq:quad}) and (\ref{eq:cubic}) together imply the $d_i$ cannot all be zero, or else there would be no rational solution for the $e_i$ charges, and then the linear constraint $D=\sum_i d_i=0$ implies at least two of the $d_i$ are non-zero.

\subsection{Classifying solutions with one neutral lepton} \label{sec:one-lepton}

As advertized in \S \ref{sec:neutrality-offpeak}, to decouple the $Z^\prime$ from FCC-ee, indeed even from old LEP-II data, means it should not couple to the electron at tree-level.
It is therefore of interest to set $e_1=0$ and study the space of solutions satisfying this condition.\footnote{To investigate solutions with three non-zero $e_i$ charges, one could apply the geometric tools pioneered in the most general SM$\times U(1)_X$ case in Refs.~\cite{Allanach:2019gwp,Allanach:2019uuu,Allanach:2020zna}. } 
The space of all such $Z^\prime$ models, satisfying our anomaly-free and electroweak-neutrality criteria, is equivalent to the set of Pythagorean triples and so admits a straightforward analytic parametrization, as follows.

We know $\sum e_i = 0$ from the linear equations, therefore if we suppose one lepton is neutral, taking $e_1=0$ wlog, then all solutions are of the form $e_2=-e_3$. The cubic then becomes $\sum_i d_i^3 = 0$, which implies by Fermat's theorem that $d_i\propto \{0,1,-1\}$ also up to permutation. Therefore, all solutions in which one right-handed lepton is neutral are of the `parity-symmetric' form, namely for each right-handed field of type $f \in \{u,d,e\}$ the charge assignment is $(f_1,f_2,f_3) = X_f (1,-1,0)$ up to permutation. 
The $X_f$ values, which we can define to all be non-negative without loss of generality, must then be chosen to satisfy the quadratic, which becomes
\begin{equation} \label{eq:pythag}
    2X_{u}^2=X_{d}^2+X_{e}^2\, .
\end{equation}
For example, one solution is 
    $X_u = X_d = X_e = 1$,    
which we discuss further below when we turn to phenomenology.
Equation \eqref{eq:pythag} can be mapped to Pythagoras' equation, by rewriting it as
\begin{equation} \label{eq:pythagX}
    X_{u}^2=\left(\frac{X_{d}-X_{e}}{2}\right)^2 + \left(\frac{X_{d}+X_{e}}{2}\right)^2\, ;
\end{equation}
since $X_{d}$ and $X_{e}$ are either both even or both odd to satisfy \eqref{eq:pythag}, we know $(X_{d}\pm X_{e})/2$ are themselves both integers. 
The complete solution (over the integers) to Pythagoras' equation has been known for over two millenia.
The set of all $Z^\prime$ models of interest can thence be parametrized in terms of two integers $p,\, q\in \Z^2$,
\begin{align} \label{eq:euclid}
    X_u = p^2 + q^2, \qquad
    X_d = p^2 + 2pq -q^2, \qquad
    X_e = q^2 + 2pq - p^2\, ,
\end{align}
from Euclid's formula. We pick out some benchmarks to study in \S \ref{sec:pheno}.

\subsection{Including right-handed neutrinos} \label{sec:neutrino}

We have found that combining the requirements of electroweak neutrality (on and off the $Z$ peak) with anomaly cancellation is very predictive for $Z^\prime$ extensions of the SM: to sum up, it implies the $Z^\prime$ couples at tree-level to up and charm quarks, at least two down-type quarks, muons, and taus, and that all inequivalent solutions up to family permutations are parametrized by a pair of integers.

One might wonder whether the above constraints can be relaxed by including up to three right-handed neutrinos, with non-universal charges $\{n_i\}$ with which one can try to `absorb' gauge anomalies.\footnote{By extending the SM by such `dark states', one can also deplete the $Z^\prime$ branching ratio into visible particles, altering the collider phenomenology ({\em e.g.} by opening up mono-jet searches). We do not consider such constraints in this paper, but constrain only the effective interactions amongst visible particles induced by the $Z^\prime$ (in \S \ref{sec:pheno}). } 
Given the linear conditions excluding the mixed anomaly with gravity, none of which notice the inclusion of SM singlets, imply that $U=D=E=0$, in order to not re-introduce the mixed anomaly with gravity we require also $N:=\sum_{i=1}^3 n_i = 0$. Then, the only modification is that the cubic becomes 
\begin{equation}
    0=\sum_i 3d_i^3+e_i^3+n_i^3\, .
\end{equation}
Since the quadratic equation is not modified, the conclusion that right-handed up and charm quarks are charged is unchanged. Phenomenologically, there is no way to decouple the di-jet constraints.

At first glance it appears slightly more encouraging to try to find solutions in which, say, all the charged leptons decouple, if we are hoping to hide from LHC Drell--Yan constraints. Setting all $e_i=0$, we now open up non-trivial solutions to the cubic in which all three $d_i$ are non-zero thanks to the inclusion of $\{n_i\}$. The quadratic constraint would then be
\begin{equation}
    4u_1^2=d_1^2+d_2^2+d_3^2\, .
\end{equation}
Considered in isolation, this equation has infinitely many solutions that can be straightforwardly parametrized in terms of four rational variables (by a generalisation of Euclid's formula), with all $d_i$ necessarily even. 
But it turns out that none of these solutions intersect the linear constraints imposed above.
To show this, our strategy is to substitute the linear constraint $D=\sum_i d_i=0$ into this quadratic to reduce the number of variables by one, and then appeal to Legendre's theorem on the ternary quadratic form.

It is convenient to perform a change of variables on the $\{d_i\}$.
Following~\cite{Allanach:2018vjg}, for instance, we define $d_{32}:=d_3-d_2$ and $\bar{d}:=-2d_1+d_2+d_3$.
Note that the linear map from $\{d_1,d_2,d_3\}$ to $\{D,d_{32},\bar{d}\}$ and its inverse are rational.
In the new variables, the quadratic becomes
$3d_{32}^2+\bar{d}^2+2D^2 = 24 u_1^2$.
In this form, it is easy to impose the linear condition, which is just $D=0$, giving
\begin{equation}
    d_{32}^2+3\left(\frac{\bar{d}}{3}\right)^2 = 2 (2u_1)^2\, .
\end{equation}
We have cast the quadratic equation in the form $ax^2+by^2=cz^2$, where the coefficients $a$, $b$ and $c$ are coprime and square-free positive integers. 
By Legendre's theorem, this does not admit any rational solutions for $\{x,y,z\}$ because the Legendre symbol $(ac|b)=(2|3)=-1$.\footnote{ Recall that, in number theory, the Legendre symbol $(a|p)$ is, given an odd prime $p$, 
defined to equal $1$ if $a$ is congruent to a (non-zero) perfect square modulo $p$, equal to $0$ if $a$ is a multiple of $p$, and equal to $-1$ otherwise. 
In the case at hand, we have that $(2|3)=-1$ because every square is congruent to 0 or 1 modulo 3.
}
Accordingly, there are no rational solutions for our original variables $\{u_1,d_i\}$ to which $\{x,y,z\}$ are related by a rational linear transformation.

We thus prove by contradiction that there exist no anomaly-free non-sterile solutions that satisfy our $Z$-pole invisibility criteria in which the charged leptons are all neutral, even allowing for arbitrarily many right-handed neutrinos with any rational charges. As before, the linear constraint $E=\sum e_i=0$ then implies at least two $e_i$ are non-zero,\footnote{The same proof can be given to show that at least two down-type quarks are also charged.} 
which implies the $Z^\prime$ must couple at tree-level to either electrons or muons. As argued above, in order to hide the $Z^\prime$ from $e^+e^-$ colliders, {\em i.e.} to make the $Z^\prime$ invisible both on and off the $Z$-peak at FCC-ee, we learn unambiguously that such a $Z^\prime$ couples at tree-level to muons as well as quarks, making it ideal to search for at the LHC.

\subsection{Compatibility with semi-simple UV completions} \label{sec:unification}

While $Z^\prime$ extensions of the SM might be motivated by many phenomenological reasons, for instance to explain various `anomalies' in flavour physics or to explain aspects of the flavour puzzle, one theoretical argument against such extensions is that they typically run counter to our desire for unification into a semi-simple gauge group. Unification remains attractive for curing the SM of the Landau pole associated with hypercharge while also explaining the observed quantization of hypercharge. Extending the SM gauge group by a second $U(1)_X$ factor near the TeV scale doubles the challenge of unifying into a semi-simple gauge group; to get an idea, note that if there were only one generation of SM fermions (or equivalently if we impose flavour universality), then only the choice $B-L$ is compatible with a semi-simple UV gauge embedding.

With three generations, there are significantly more options for gauged $U(1)_X$ extensions of the SM that have semi-simple embeddings: the options were completely classified in Ref.~\cite{Davighi:2022dyq}, assuming up to three RH neutrinos, by considering all possible Lie algebra embeddings of $\su(3)\oplus \su(2)\oplus \mathfrak{a} \hookrightarrow \g \hookrightarrow \su(48)$ consistent with the SM fermion representation (where $\mathfrak{a}$ denotes an abelian Lie algebra containing hypercharge, $\mathfrak{g}$ is the semi-simple Lie algebra we wish to embed in, and $\su(48)$ is the maximal SM flavour symmetry with all interactions turned off), and then checking for (local and global) anomaly cancellation.

We now ask which of these extensions satisfy our additional criteria for not running at 1-loop into the EWPOs. The key observation that makes this in fact very simple is that electroweak neutrality required all LH fermions be neutral, as explained in \S \ref{sec:neutrality}. Given that, we can infer that the only anomaly-free $U(1)_X$ choices with semi-simple embeddings are of the form~\cite{Davighi:2022dyq}
\begin{equation}
    u_i=d_i=e_i=n_i=(a,b,-a-b) 
\end{equation}
up to permutations within families.
The condition from $y_t$ and $g_3$ running, which imposed $u_3=0$, then restricts us to a unique solution up to an overall normalisation
\begin{equation}
    u_i=d_i=e_i=n_i=(1,-1,0)\, ,
\end{equation}
where again we are free to permute the family indices for $d$, $e$, and $n$, which leads to different phenomenology (see \S \ref{sec:pheno}). So, strikingly, there is a {\em unique} anomaly-free $U(1)_X$ extension of the SM that doesn't run into the $Z$-pole observables {\em and} which has a semi-simple UV completion. We further emphasize that including (at least two) right-handed neutrinos is now essential to reveal this possibility.\footnote{The inclusion of right-handed neutrinos can also change the phenomenological signatures of such an anomaly-free $Z^\prime$, for instance by opening up invisible decay channels if the right-handed neutrinos are light enough, as studied in {\em e.g.}~\cite{Das:2017flq,Das:2022rbl}. }

This $U(1)_X$ charge assignment embeds inside an $Sp(4)_R$ `electroweak flavour unification' symmetry, studied in~\cite{Davighi:2022fer,Davighi:2022vpl,Davighi:2022bqf}, in which right-handed up- and down-type fields (both quarks and leptons) of two generations are unified into ${\bf 4}$-dimensional multiplets which must be implemented alongside quark-lepton unification via the $SU(4)$ of Pati and Salam~\cite{Pati:1974yy}.\footnote{
The results of~\cite{Davighi:2022dyq}, for instance via the \texttt{TestYourOwnCharges.nb} Mathematica notebook provided \href{https://arxiv.org/src/2206.11271v2/anc}{here}, can be used to deduce other semi-simple UV completions for this $Z^\prime$ model, all of which feature either some electroweak flavour unification for right-handed fields, or feature $SO(10)$ unification for all three families.
} The $Sp(4)_R$ symmetry could be further embedded inside an $Sp(6)_R$; a candidate semi-simple embedding is then given by the group $SU(4) \times Sp(6)_L \times Sp(6)_R$, which can break down to the SM via our SM $\times U(1)_X$ extension.
An alternative embedding of the right-handed $U(1)_X$ is inside the group $SO(6)_R$. All of these options are free of local and global gauge anomalies.

This unique charge assignment forms one of our benchmarks when we come to study the phenomenology in \S \ref{sec:pheno}; from the phenomenological perspective it is an instructive benchmark because of the equal couplings to quarks and leptons.

\subsection{Comments regarding quark and lepton flavour structures} \label{sec:flavour-structure}

All the chiral extensions of the SM considered above do not permit the full Yukawa coupling matrices at the renormalisable level. As mentioned in \S \ref{sec:neutrality}, the conditions $H=q_3=u_3$ do, at least, guarantee the top quark Yukawa coupling is consistent with the SM$\times U(1)_X$ gauge symmetry for all models we consider. Beyond that, the gauge non-invariance of many elements of the Yukawa matrices means they must be populated by higher-dimension operators, which requires further states in the UV beyond just the $U(1)_X$ gauge field and the scalar field that breaks it, such as vector-like fermions. The Yukawa matrices in these models are therefore expected to possess hierarchical structure, with (at least) an order-1 top quark Yukawa coupling.
Since our goal in this paper is not to address the flavour puzzle, but rather to explore the phenomenological complementarity of different experiments in probing anomaly-free $Z^\prime$ models, 
we are content to make only a few comments concerning flavour textures.

The parity-symmetric solutions in \S \ref{sec:one-lepton}, for instance, admit a full column of each Yukawa matrix $Y_{u,d,e}$ at the renormalisable level. 
For up-type quarks this must be the third column, and for down-type quarks it is natural to assume the same since then $y_b$ is also unsuppressed. But for leptons, we already took $e_1=0$ for phenomenological reasons. Such $U(1)_X$ symmetries therefore admit the first column of the lepton Yukawa but not the second or third. A natural implementation of these model presumably entails a different mechanism deeper in the UV (such as an additional gauge symmetry) that results in suppression of $y_e$.

Finally, it is worth remarking that the quark Yukawa textures na\"ively predicted by gauging $U(1)_X$ are consistent with small mixing angles amongst the right-handed quarks, since, if we take $d_3=0$ to be the neutral right-handed down quark flavour, then we generally expect
    $Y_{u,d,e} \sim
    \begin{psmallmatrix}
        \delta & \epsilon & 1 \\
        \delta & \epsilon & 1 \\
        \delta & \epsilon & 1
    \end{psmallmatrix}$,
where $\epsilon$ and $\delta$ are small parameters related to the breaking of $U(1)_X$.\footnote{For instance, in a Froggatt--Nielsen-like setup~\cite{Froggatt:1978nt}, these take the form $(v_X/M)^{n}$ where $M$ is the mass of a heavy fermion and $n$ is some integer.} With such a texture, right-handed rotations would have the form 
    $V_R \sim 
    \begin{psmallmatrix}
        1&\delta&\delta\epsilon\\
        \cdot &1&\epsilon\\
        \cdot&\cdot&1
    \end{psmallmatrix}$,
which further justifies our expectation that models in this class can have small flavour-violation and thus be more-or-less flavour neutral.

\section{Phenomenology on and off the $Z$ pole} \label{sec:pheno}

The set of $Z^\prime$ models that we have identified in the previous Sections, which are anomaly-free and do not run via 1-loop RGEs into $Z$-pole observables, all feature couplings to right-handed quarks and leptons, but nothing else. When the $Z^\prime$ is integrated out at tree-level, these models give rise to a set of flavour-conserving 4-fermion SMEFT operators with `RR' chirality structure, which we can categorise into $4\ell$, $2q2\ell$, and $4q$ operators. The former are strongly constrained if there is a coupling to electrons, by cross-section measurements at electron-positron colliders (\S \ref{sec:4l}). Seeking invisibility at lepton colliders, this motivates coupling the $Z^\prime$ only to muon and tau leptons. The semi-leptonic operators, being flavour-conserving, are then best constrained by Drell--Yan data from the LHC (\S \ref{sec:2q2l}), which are reasonably strong for all lepton flavours in the final state. To hide from Drell--Yan $pp\to\ell\ell$ also, we can try to further restrict to $Z^\prime$ models in which the lepton charges are numerically very small. But even in this most invisible case, the $4q$ operators can be constrained from di-jet resonance searches at the LHC (\S \ref{sec:4q}).

We neglect the effects of renormalisation group (RG) running of the SMEFT operators from the matching scale $\Lambda$ down to the relevant experimental energy scale. 
As explained in \S \ref{sec:RG}, this enters only at higher-order in the Wilson coefficients and/or is suppressed by light Yukawa couplings. For the models we focus on with $e_1=0$ and $d_3=0$, the leading RG effects are suppressed by $y_c^2 \sim 10^{-5}$ in addition to the loop factor and heavy mass scale, and are thus totally negligible.

\subsection{Benchmark Models}


To assess the strength of the various bounds, it is instructive to define two benchmark (BM) solutions, both within the class identified in \S \ref{sec:one-lepton} for which electrons are neutral, as follows:
\begin{align} \label{eq:BMs}
    &\text{Benchmark 1:~} (X_e=X_{u,d})   &&(X_e,X_u,X_d)\propto(1,1,1)\, , \\
    &\text{Benchmark 2:~} (X_e\ll X_{u,d})  &&(X_e,X_u,X_d)\propto(1,29,41)\, .    
\end{align}
The first BM model has equal charges for quarks and leptons, and gives a suitable BM for assessing the strength of the Drell--Yan constraints. 
As discussed in \S \ref{sec:unification}, this benchmark is special from the theoretical point of view because it is the only possible $Z^\prime$ extension in our class that has a semi-simple UV completion.
The second BM has a numerically very small lepton charge,\footnote{To obtain BM2, the Pythagorean triple solving (\ref{eq:pythagX}) corresponds to a right-angled triangle whose perpendicular sides are of near-equal length, in this case the $(20,21,29)$ triangle.}
for which the Drell--Yan constraints also become subleading, leaving us with LHC dijet searches as the leading constraints.

We point out that a benchmark in which $X_e \gg X_{u,d}$, which we might hope to decouple from the LHC entirely and make truly invisible by coupling only to RH taus and muons, is {\em not} possible: this is because it is $X_u$ that appears on the RHS of the quadratic ACC, which recall was $X_d^2+X_e^2=2X_u^2$, and so $X_u$ cannot be made small. In particular, we have the strict inequality
\begin{equation}
    X_e \leq \sqrt{2} X_u\, .
\end{equation}
Thus, the benchmarks in \eqref{eq:BMs} cover the two main limiting cases available to us, in which leptons either have comparable charge to quarks, or they have far smaller charge.

\subsection{Finite 1-loop contributions on the $Z$-pole} \label{sec:Zpole}

While the EWPOs receive vanishing contributions from these models at the leading log approximation to the 1-loop running, there are nonetheless finite 
({\em i.e.} non logarithmically divergent) parts of the 1-loop diagrams in the full UV theory that should enter into the computation of low-energy observables (see {\em e.g.}~\cite{Gargalionis:2024jaw} for FCC-ee studies including similar effects).
Precisely because there is no effect up to leading log, it is important to estimate the size of these effects before we conclude that $Z$-pole and $W$-pole observables do not strongly constrain these models, which we do here.

For our models, the vanishing dependence of loop-induced effects on the top Yukawa, which note is the largest SM coupling of relevance, holds not just at leading log but for the full 1-loop contribution, simply by vanishing of the diagrams in~\eqref{eq:yt-loop}.
There are, however, non-vanishing finite parts dependent on the electroweak gauge couplings. This cannot be seen from the SMEFT loops in {\em e.g.}~\eqref{eq:feynman} assuming tree-level matching at the high scale as we have done, but can be seen from inspection of the 1-loop diagrams in the UV theory.\footnote{I am grateful to Ayres Freitas for highlighting this point, and to Guilherme Guedes, Matthew McCullough, and especially Lukas Allwicher for helpful discussions concerning these finite 1-loop matching contributions to the $Z$-pole observables. } 
In particular, the modification to the $Z$-boson couplings to fermions, for which the SMEFT RGE contributions are captured by the diagram in \eqref{eq:feynman}, receives contributions from two types of diagram in the full theory:
\begin{equation} \label{eq:finite}
        \begin{tikzpicture}[baseline=-0.5ex]
        \begin{feynman}
        \vertex (a) {$Z_\mu$};
        \vertex [right=0.6in of a] (b);
        \vertex [above right=0.6in of b] (c);
        \vertex [below right=0.6in of b] (d);
        \node at (c) [circle,fill,inner sep=1.2pt];
        \node at (d) [circle,fill,inner sep=1.2pt];
        \vertex [above right=0.3in of c] (c1) {$f_i$};
        \vertex [below right=0.3in of d] (d1) {$f_i$};
        \diagram{
        (a) -- [photon](b), 
        (b) -- [edge label=$\,\,\,f_i$] (c) -- (c1),
        (c) -- [photon, blue, edge label=$Z^\prime$] (d),
        (b) -- [edge label'=$\,\,\,f_i$] (d) -- (d1),
        };
        \end{feynman}
        \end{tikzpicture}
        \quad + \quad \sum_{f_k^\prime}\,\,
        \begin{tikzpicture}[baseline=-0.5ex]
        \begin{feynman}
        \vertex (a) {$Z_\mu$};
        \vertex [right=0.6in of a] (b);
        \vertex [right=0.5in of b] (c);
        \node at (c) [circle,fill,inner sep=1.2pt];        
        \vertex [right=0.4in of c] (d);
        \node at (d) [circle,fill,inner sep=1.2pt];        
        \vertex [above right=0.4in of d] (e) {$f_i$};
        \vertex [below right=0.4in of d] (f) {$f_i$};
        \diagram{
        (a) -- [photon](b),
        (b) -- [half left, edge label=$f^\prime_k$] (c) -- [half left] (b),
        (c) -- [photon, blue, edge label=$Z^\prime$] (d),
        (d) -- (e),
        (d) -- (f),
        };
        \end{feynman}
        \end{tikzpicture}        
\end{equation}
The heavy $Z^\prime$ particle is highlighted in blue.
The second diagram gives no finite 1-loop matching corrections, because the loop integral is entirely captured by the corresponding loop in the SMEFT diagram~\eqref{eq:4f_renorm}. 
But the first diagram gives a non-zero 1-loop matching contribution to the SMEFT coefficients $C_{Hf}^{ii}$, proportional to $g_1^2$. 
By computing the finite parts of both the full UV diagram and the SMEFT diagram, and taking the difference between the leading term in the resulting $1/M_X^2$ expansion, we extract 
\begin{equation}
    C_{Hf}^{\text{1-loop, finite}} = -\frac{7Y_H}{9}\frac{g_1^2}{16\pi^2} \frac{g_X^2}{M_X^2} \begin{cases}
         Y_e X_{e_i}^2 \qquad &f=e_i ,\\
         Y_u X_{u_i}^2 \qquad &f=u_i ,\\
         Y_d X_{d_i}^2 \qquad &f=d_i ,      
    \end{cases}
\end{equation}
where $Y_a$ denotes the hypercharge of SM field $a$. We input these 1-loop matched WC values into a projected FCC-ee electroweak fit to estimate the tera-$Z$ sensitivity to the benchmark `invisible' $Z^\prime$ models we have derived, finding (see Fig.~\ref{fig:BAR}):
\begin{equation}
    \frac{M_X}{g_X} \gtrsim \begin{cases}
        850 \text{~GeV}, \quad & \text{(BM1)} \\
        250 \text{~GeV}, \quad & \text{(BM2)}        
    \end{cases}
\end{equation}
Thus, the finite 1-loop matching contributions are sufficiently small that a tera-$Z$ run at FCC-ee does not constrain these models up to the TeV scale. The substantially weaker bound for BM2 comes from the fact that only the $Z$ couplings to (non-$b$) jets is modified.

\subsection{$4\ell$: $e^+e^-\to \ell\ell$ } \label{sec:4l}

If the $Z^\prime$ couples at tree-level to electrons, there are bounds coming from measurements of the $e^+e^- \to \ell_i \ell_i$ cross-section (and forward-back asymmetry measurements) at electron-positron colliders. There are strong constraints already coming from the LEP-II measurements, which we calculate here for the $Z^\prime$ models of interest.
The reach of FCC-ee will far exceed the LEP-II bounds, thanks to the proposed set of FCC-ee runs above the $Z$-pole.

The LEP-II constraints have been computed recently for $Z^\prime$ models in {\em e.g.} Refs.~\cite{Greljo:2022jac,Allanach:2023uxz}, based on the $\chi^2$ likelihood computed in~\cite{Falkowski:2015krw}. The observables that go into computing the likelihood are the cross-sections 
    $\sigma_{\ell\ell}:=\sigma(e^+e^- \to \ell^+\ell^-)$ for $\ell \in \{e,\mu,\tau\}$,
and the forward-backward asymmetry observables (defined as the forward cross-section minus the backward cross-section), for $\mu^+\mu^-$ and $\tau^+\tau^-$ final states.
The relevant Wilson coefficients turned on in our $Z^\prime$ models are $O_{ee}^{11jj}$, where $j$ can be any lepton flavour; without turning on the electron coupling, there are no bounds at leading order. 
It is therefore instructive to hone in on the `model-independent' constraint that is unavoidable if we turn on only the electron coupling, which is the bound from $\sigma_{ee}$ on $C_{ee}^{1111}=-g_X^2 X_e^2/(2M_X^2)$, where $X_e$ is the $e_1$ charge under $U(1)_X$. The $95\%$ C.L. constraint, obtained from the $\chi^2-\chi^2_{\text{min}}=3.84$ contour (given we have a single independent model-parameter),\footnote{We are very grateful to Lukas Allwicher for sharing code for the LEP-II $\chi^2$ likelihood that we used in this Section, and for related helpful discussions.
} is
\begin{equation} \label{eq:4lbound}
    \frac{M_X}{X_e g_X} \gtrsim 5.2 \text{~TeV}\, .
\end{equation}
We remark that, if the Wilson coefficient $C_{ee}^{1111}$ had the opposite sign, we find the bound would be $2\div 3$ times weaker than this ({\em i.e.} 2 TeV), which reflects an underlying mild tension in the $e^+e^-\to e^+e^-$ data from LEP-II. This is in agreement with the results shown in~\cite[Fig. 1]{Falkowski:2015krw}.

Of course, we already learnt that at least the tau or muon charge must also be turned on to satisfy the anomaly cancellation criteria, which only strengthens the bounds. For example, consider a `parity symmetric' solution of the kind discussed in \S \ref{sec:one-lepton}, with electron and muon charged equal and oppositely (for which the lepton Yukawa would mimic those for the quarks, {\em i.e.} with $y_\tau$ being renormalisable). The SMEFT Lagrangian is
$\cL = -\frac{g_X^2}{M_X^2}X_e^2 (\bar{e}_R \gamma_\mu e_R - \bar{\mu}_R \gamma_\mu \mu_R)^2$.
Including also $\sigma_{\mu\mu}$ and the muon forward-backward asymmetry LEP-II measurement in the $\chi^2$ likelihood, the bound increases a little to
\begin{equation}
    \frac{M_X}{X_e g_X} \gtrsim 5.9 \text{~TeV}\, .
\end{equation}
If one considered the option with $e_R$ and $\tau_R$ charged, and $\mu_R$ neutral, including this time the $\sigma_{\tau\tau}$ and $\tau$ FB asymmetry observables, then the bound increases with respect to (\ref{eq:4lbound}) by only 40 GeV or so.

All these bounds on $M_X/X_e g_X$ would leap forward in reach with FCC-ee, thanks to its planned sequence of high-statistics runs above the $Z$-peak.\footnote{The expected sensitivity to $Z^\prime$ scenarios from $e^+e^- \to f\bar{f}$ at other proposed future lepton colliders, such as CLiC, ILC, and CEPC, was studied in~\cite{Das:2021esm}. } 
As recently shown in Ref.~\cite{Greljo:2024ytg}, the expected increase in reach, for example for the $C_{ee}^{1111}$ Wilson coefficient that is most important here, is by nearly an order of magnitude.
In fact, given we know that the $Z^\prime$ must couple to at least tau or muon in addition to electrons, the strongest relevant bounds from~\cite{Greljo:2024ytg} are on the (necessarily RR chirality) operators $O_{ee}^{11,22}$ and $O_{ee}^{11,33}$, which reach up to\footnote{
We also point out that lepton flavour universality violation (LFUV) is generically predicted by these models; to recap, it is forced upon us by anomaly cancellation, which implies both that $\sum_i e_i = 0$ and that at least one $e_i$ is non-zero. FCC-ee provides new opportunities for LFUV tests, such as measuring the ratio
\begin{equation}
    R_{\tau/\mu} = \frac{\sigma(e^+e^- \to \tau^+\tau^-)}{\sigma(e^+e^- \to \mu^+\mu^-)}\, ,
\end{equation}
as discussed in~\cite{Greljo:2024ytg}.
Such LFUV ratios would provide competitive probes of these $Z^\prime$ scenarios, although they are not independent from the more-typical `$R_l$' ratios that were measured at LEP. \label{foot:LFUV}
}
\begin{equation}
    \frac{M_X}{g_X} \gtrsim 30 \text{~TeV} \qquad \text{(FCC-ee above $Z$-peak)}\, ,
\end{equation}
which actually well exceeds the scale of $7$ TeV (in the case of flavour-universal couplings to SM fermions) that is probed due to 1-loop running into the $Z$-pole observables~\cite{Allwicher:2024sso}. 
Therefore, as anticipated, we should discard this option in our hunt for a $Z^\prime$ that is invisible to FCC-ee.

\subsection{$2q2\ell$: $pp \to \ell\ell$} \label{sec:2q2l}

Even if we decouple the $Z^\prime$ from electrons, it must then couple to both $\mu_R$ and $\tau_R$, which leads to important bounds from Drell--Yan $pp \to \mu\mu$ and $pp\to \tau\tau$ measurements at the LHC. Here we use the \texttt{HighPT} package~\cite{Allwicher:2022mcg} to estimate the current strength of these constraints. The underlying ATLAS and CMS searches~\cite{ATLAS:2020zms,CMS:2021ctt} that go into our likelihood use $139 \text{~fb}^{-1}$ of LHC data. A similar implementation of the bounds from Drell--Yan can be performed using the \texttt{smelli} software~\cite{Aebischer:2018iyb,Stangl:2020lbh,Greljo:2022jac}.

Here we focus on the solutions found in \S \ref{sec:one-lepton} in which the electron is neutral. Even though all charges are specified by two integer parameters $p$ and $q$ via \eqref{eq:euclid}, it is transparent to write things in terms of $X_{u,d,e}$ for now.
Integrating out the $Z^\prime$ gives the following tree-level dimension-6 semi-leptonic SMEFT operators
\begin{align} \label{eq:L2q2l}
    \cL = &\mp\frac{g_X^2}{M_X^2}X_{e}X_u (\bar{\tau}_R \gamma_\mu \tau_R - \bar{\mu}_R \gamma_\mu \mu_R)(\bar{u}_R\gamma^\mu u_R - \bar{c}_R \gamma^\mu c_R) \\
    &\mp\frac{g_X^2}{M_X^2}X_e X_d (\bar{\tau}_R \gamma_\mu \tau_R - \bar{\mu}_R \gamma_\mu \mu_R)(\bar{d}_R\gamma^\mu d_R - \bar{s}_R\gamma^\mu s_R)
\end{align}
The sign ambiguities out the front correspond to permuting the signs of $u_i$ or $d_i$ charges, keeping the $e_i$ permutation fixed. 

To get a feel for the bounds, it is helpful to consider benchmark 1 (BM1) defined in \eqref{eq:BMs}.
Given present data, there is a significant dependence of the bound that we extract on the possible sign choices in \eqref{eq:L2q2l}, which trace back to fluctuations in the data giving preference for non-zero values of the Wilson coefficients turned on by our $Z^\prime$ model.
We obtain
\begin{equation} \label{eq:DY4bounds}
    \frac{M_X}{g_X} \gtrsim 
    \begin{cases}
        6.7 \text{~TeV,} \qquad  &C_{eu}^{2211} > 0,\, C_{ed}^{2211} >0 \\
        13.8 \text{~TeV,} \qquad  &C_{eu}^{2211} < 0,\, C_{ed}^{2211} >0 \\
        9.2 \text{~TeV,} \qquad  &C_{eu}^{2211} > 0,\, C_{ed}^{2211} <0 \\
        11.3 \text{~TeV,} \qquad  &C_{eu}^{2211} < 0,\, C_{ed}^{2211} < 0
    \end{cases}
\end{equation}
We thus learn that, for equitable quark and lepton charges, the bounds from LHC Drell--Yan $pp \to \ell\ell$ data are strong, roughly excluding the $Z^\prime$ up to scales of 10 TeV or so.

These bounds, which are already probing very high scales due to the order-1 coupling of the $Z^\prime$ to both valence quarks and to charged leptons, will continue to strengthen as the LHC analyses more data, albeit slowly since the statistical uncertainty on the measured event rates goes down with the square root of the gain in luminosity.
By the end of the High Luminosity LHC upgrade, assuming an integrated luminosity of $3\text{~ab}^{-1}$ of $pp$ collision have been accrued, and that the predicted Standard Model
background rate is measured in all bins of the high-$p_T$ Drell--Yan distributions, we project the bound to be
\begin{equation}
    \frac{M_X}{g_X} \gtrsim 22 \text{~TeV} \qquad \text{(High-Luminosity LHC)}\, ,
\end{equation}
which we obtained again using the \texttt{HighPT} package~\cite{Allwicher:2022mcg}, with the bound being largely driven by the di-muon search, and where we average over the bounds one would obtain with the four different choices for the relative signs of charges, as in \eqref{eq:DY4bounds}.

Even though we discarded the options in which electrons are charged in our hunt for an invisible $Z^\prime$ due to the very strong constraints one would obtain at FCC-ee above the $Z$-peak, we can give the corresponding HL-LHC projections in the other two scenarios for completeness, that is, in which electron and tau are charged and in which electron and muon are charged. We find
\begin{align}
    \frac{M_X}{g_X} \gtrsim 19 \text{~TeV} \qquad \text{for BM1 ($e\tau$)}, \\
    \frac{M_X}{g_X} \gtrsim 24 \text{~TeV} \qquad \text{for BM1 ($e\mu$)},
\end{align}
where again we averaged over the various sign choices.
Unsurprisingly, the option in which both light leptons are charged is expected to get the strongest constraint from Drell--Yan.

\subsection{$4q$: $pp\to jj$} \label{sec:4q}

The closest we get to an invisible $Z^\prime$ is one in which the lepton charge, while necessarily non-zero, is very small compared to the quark charges, as provided for instance by BM2 in \eqref{eq:BMs}.
If we normalise the gauge coupling so that the largest charge is unity, then the lepton charge is a few percent, and the semi-leptonic signal in Drell--Yan is much suppressed. Of course, the physical, normalisation-independent quantities to compare between different $Z^\prime$ models are not charges or gauge couplings, but rather the ratios between different Wilson coefficients generated by the $Z^\prime$. In particular, for BM2 we have the scaling
\begin{equation}
    |C_{4q}| \sim \frac{X_{u,d}}{X_e} |C_{2q2l}| \sim \left( \frac{X_{u,d}}{X_e}\right)^2 |C_{4l}|\, , 
\end{equation}
with $X_u/X_e=29$ and $X_d/X_e=41$, meaning that 4-quark operator coefficients are over an order-of-magnitude larger than the semi-leptonic operator coefficients studied in \S \ref{sec:2q2l}. 
We find that, for these BMs, the projected HL-LHC constraints from $pp\to\ell\ell$ drop to about 
\begin{equation} \label{eq:BM2ppll}
    M_X/g_X \gtrsim 2.5 \text{~TeV}\, ,    \qquad \text{ (BM2)}\, , 
\end{equation}
an order-of-magnitude reduction in reach compared to BM1.

These BM2 $Z^\prime$ bosons will also give unsuppressed contributions to $q\bar{q} \to q\bar{q}$ Drell--Yan, which are an unavoidable prediction of all the `electroweakly neutral' $Z^\prime$ theories we classified. This is probed by di-jet measurements at the LHC, the results of which were helpfully recast for narrow-width BSM mediators in generic SM gauge representations in Ref.~\cite{Bordone:2021cca} based on ATLAS~\cite{ATLAS:2018qto,ATLAS:2019fgd} and CMS~\cite{CMS:2018mgb,CMS:2019gwf} resonant searches. EFT-based non-resonant analyses for di-jets were performed in~\cite{ATLAS:2017eqx,CMS:2018ucw}.

Note that for this type of benchmark model, {\em i.e.} featuring $|X_e| \ll |X_{u,d}|$, the anomaly cancellation condition $2X_u^2= X_d^2+X_e^2 \approx X_d^2$ directly implies
    $|X_d| \approx \sqrt{2} |X_u|$
which, in light of the results of~\cite{Bordone:2021cca}, makes the $dd$ and $uu$ channels roughly equally strong. We can thus characterize BM2 via the approximate set of charges
\begin{equation} \label{eq:BM2_norm}
    X_d=1, \qquad X_u \approx \frac{1}{\sqrt{2}}, \qquad X_e \ll 1\, ,
\end{equation}
where we choose a normalisation for the gauge coupling $g_X$ in which the largest charge, that of down-type quarks, is unity.

Our goal in this Subsection is to use the results of~\cite{Bordone:2021cca,CMS:2018ucw} to crudely estimate conservative bounds on such a $Z^\prime$ model coming from dijet resonant and non-resonant searches. The four resonant searches searches cover different overlapping regions of the kinematical range $m_{jj}\in (0.05,5)$ TeV.
We emphasize that the bound we estimate from the resonant searches, following~\cite{Bordone:2021cca}, is qualitatively different from those for $4l$ and $2q2l$ that we studied in previous Subsections, which were obtained by integrating out the heavy $Z^\prime$ and matching onto the 4-fermion SMEFT operators to bound the effective scale $\Lambda_X = M_X/g_X$.  For the narrow resonance di-jet search, the result is not a mass-independent bound on the scale as in \S\S \ref{sec:4l} and \ref{sec:2q2l}, but a bound that only holds within the kinematical range $M_X \leq 5$ TeV -- indeed, as we next show, we cannot reliably get a bound even for this full range.

Constraining such a BSM non-resonant contribution that lifts the tails of the invariant mass distribution is a qualitatively different challenge for dijet distributions than for dileptons (\S \ref{sec:2q2l}), both experimentally and theoretically (see {\em e.g.}~\cite{ATLAS:2017eqx,Alioli:2017jdo,CMS:2018ucw}, where angular distributions of the dijet system are leveraged).
While it is beyond the scope of this work to recast the experimental analyses to our model case, we nonetheless compare the resonant search bounds with an EFT-derived bound from~\cite{CMS:2018ucw} in the case that most closely matches our model, which gives a similar ballpark bound.

\paragraph{Narrow dijet resonance search.}

To reliably interpret the resonance search to put a bound on our $Z^\prime$, the latter should be a narrow resonance with $\Gamma_X/M_X \lesssim 0.1$. We can compute the width by summing over the partial widths to $q\bar{q}$ bilinears:
\begin{equation}
    \frac{\Gamma_X}{M_X} = \frac{g_X^2}{8\pi} (2|X_u|^2+2|X_d|^2 + \dots) \approx \frac{3g_X^2}{8\pi}\, ,
\end{equation}
where the $\dots$ indicates the partial width to leptons which, for this BM2, is negligible. The narrow width condition thus puts an upper bound on the gauge coupling:
\begin{equation}
    g_X < \sqrt{\frac{4\pi \times 0.1}{|X_u|^2+|X_d|^2}} \approx 0.9\, \qquad \text{(BM2)},
\end{equation}
given the choice of charge normalisation in \eqref{eq:BM2_norm}. The dijet resonance search constraints from~\cite{Bordone:2021cca} can only be reliably used when this condition on $g_X$ is satisfied.
Because we are interpreting a resonance search, we do not integrate out the $Z^\prime$ and match to SMEFT. 
Rather, the relevant Lagrangian is described by the following couplings of the $Z^\prime$ to quarks,
\begin{equation}
    L = g_X X_u (\overline{u} \gamma^\mu P_R u - \overline{c} \gamma^\mu P_R c ) X_\mu 
    + g_X X_d (\overline{d} \gamma^\mu P_R d - \overline{s} \gamma^\mu P_R s ) X_\mu \, .
\end{equation}
Given the equal-size first and second generation quarks, and the fact that the parton distribution functions (PDFs) are much suppressed for the second generation with respect to the first, 
we can drop the charm and strange quark contributions from now on.

From digitising the plot in~\cite{Bordone:2021cca}, 
the perturbativity limit $g_X \lesssim 0.9$ (from $\Gamma_X/M_X < 1$) implies that we can use the constraint on the coupling to $d\bar{d}$, which is $X_d g_X \approx g_X$, out to masses $M_X \leq 4.2$ TeV. Likewise in the $u\bar{u}$ channel, the bound can be used up to slightly higher $Z^\prime$ masses $M_X \leq 4.4$ TeV, where we use the fact that $X_u g_X \approx g_X/\sqrt{2}$ from \eqref{eq:BM2_norm}. 
Then at this end-point of validity, we can exclude this value of the mass and coupling; for that mass, all couplings $g_X > 0.9$ are then robustly excluded. 
To compare with our other EFT-derived constraints, which all depend on a scale $\sim M/(gX)$ rather than a mass and coupling individually, we can phrase the bound as
\begin{equation}
    \frac{M_X}{g_X} \geq 4.9 \text{~TeV}, \qquad \text{valid for } M_X \leq 4.4 \text{~TeV}\, ,
\end{equation}
which surpasses the reach \eqref{eq:BM2ppll} from $pp\to\ell\ell$.

We admit that this is a very crude estimate of the bound on our $Z^\prime$ coming from di-jet, but it is equally very conservative: we use only the resonance search, and we use it only in the region where the width over mass is $<0.1$. This is already sufficient to demonstrate that the bound is strong, at the level on 5 TeV on effective scale, meaning this $Z^\prime$ is far from being invisible at the LHC. The bound is driven by the large coupling to up quarks, which we showed is necessary for all $Z^\prime$ models that do not run into the $Z$-pole, which clearly emphasizes the perfect complementarity between the LHC and a future FCC-ee machine in probing {\em any} $Z^\prime$ extension of the SM that is not sterile.

\paragraph{4-quark operator bound from dijet angular distributions.}

In Ref.~\cite{CMS:2018ucw} the CMS collaboration perform an analysis of the angular distributions in dijet final states to obtain bounds on 4-quark effective operators using 35.9 $\text{fb}^{-1}$ of 13 TeV data, for which the interpretation is not restricted by a kinematical upper limit on the mass. The scenario they consider that is closest to our family of $Z^\prime$ models is one with flavour-universal couplings to only right-handed quarks (corresponding to the Wilson coefficient $\eta_{RR}$ in the experimental analysis~\cite{CMS:2018ucw}). Given BM2 has a similar strength coupling to $u_R$ and $d_R$, we are not too far from this flavour-universality limit. 
Translating to our SMEFT Lagrangian, and taking both $X_u$ and $X_d$ to be approximated by the smaller of the two in our benchmark in order to get a conservative estimate of the bound,
we infer the following EFT-derived  bound on the $Z^\prime$ scale
\begin{equation} \label{eq:4qEFTbound}
    \frac{M_X}{g_X} \geq 
        3.5 \text{~TeV} \qquad \text{valid for all $M_X$ (up to perturbativity)}\, .
\end{equation}
We reiterate that this bound would correspond to a $Z^\prime$ with couplings $X_u=X_d=1/\sqrt{2}$, whereas for our `leptophobic' anomaly-free class of models (realised by BM2 in \eqref{eq:BM2_norm}) we have $X_u\approx 1/\sqrt{2}$ and a larger $X_d=1$, so if we performed a proper recast for BM2 one would expect the EFT bound to be somewhat stronger than \eqref{eq:4qEFTbound} thanks to the enhanced down-quark coupling.

\subsection{Comment on low-energy probes of LFUV}

For completeness, we make some comments about low-energy flavour probes before moving to our summary. Recall from footnote~\ref{foot:LFUV} that our $Z^\prime$ models necessarily feature LFUV, with all three RH lepton charges being different. We there mentioned the possibility of probing this in $e^+e^-$ colliders, from measuring the LFUV ratio $R_{\tau/\mu}$ (indeed, there would already be bounds coming from measurements of $R_{\tau}$ and $R_\mu$ at LEP). One might further wonder whether there are {\em low-energy} probes of this LFUV, for instance in $B$-meson decays and the class of $R_H:=BR(B\to H \mu\mu)/BR(B\to Hee)$ measurements, where $B$ is a $b$-hadron and $H$ is some other hadron.
These observables are extremely sensitive probes of new physics due to the change of quark flavour, typically $b\to s$ (as in $R_{K^{(\ast)}}$~\cite{LHCb:2022vje}, $R_{K_s}$ and $R_{K^{\ast +}}$~\cite{LHCb:2021lvy}, the very recent $R_\phi$ measurement which tests LFUV in $B_s^0\to \phi \ell^+\ell^-$~\cite{LHCb:2024rto}), for which the SM contribution is extremely suppressed. 

The heavy $Z^\prime$ bosons under consideration can significantly impact these observables if it connects a flavour-violating $bs$ vertex to a LFUV leptonic current. For left-handed $bs$ currents, the $Z^\prime$ has no direct coupling to left-handed fields and so the BSM contribution goes through the same penguin loop as for the SM process, and is a negligible correction. There is, however, a right-handed contribution at tree-level that violates LFU. 
This will be linear in the right-handed mixing angles between $b_R$ and $s_R$ which, as we discussed in \S \ref{sec:neutrality}, are expected to be small in these models and can be consistently taken to zero to decouple the corresponding bounds. 

It is perhaps worth noting that there are large, tree-level contributions to quark flavour {\em conserving} LFU tests. 
While high-precision measurements can here be made,
for instance in $J/\psi$ decays (with $\Gamma_{J/\psi \to e^+e^-}/\Gamma_{J/\psi \to \mu^+\mu^-}$ measured to be unity up to $1\%$ precision~\cite{KEDR:2013dpd} in the KEDR experiment~\cite{Anashin:2013twa}), these decays occur at tree-level in the SM and via photon exchange, and so the $Z^\prime$ will give a relative shift of order $g_X^2 M_{J/\psi}^2/e^2 M_{X}^2$ which is completely undetectable for the multi-TeV scale $Z^\prime$ mass that is forced upon us by the LHC dijet constraints alone.

Lastly, there are no LFUV effects in tau decays, because we only have neutral current LFUV but no concomitant LFV, which means that any process necessarily features an even number of taus and so cannot affect $\tau$ decays.

\subsection{Comparison of benchmark models}

\begin{figure}
    \centering
    \includegraphics[width=0.88\textwidth]{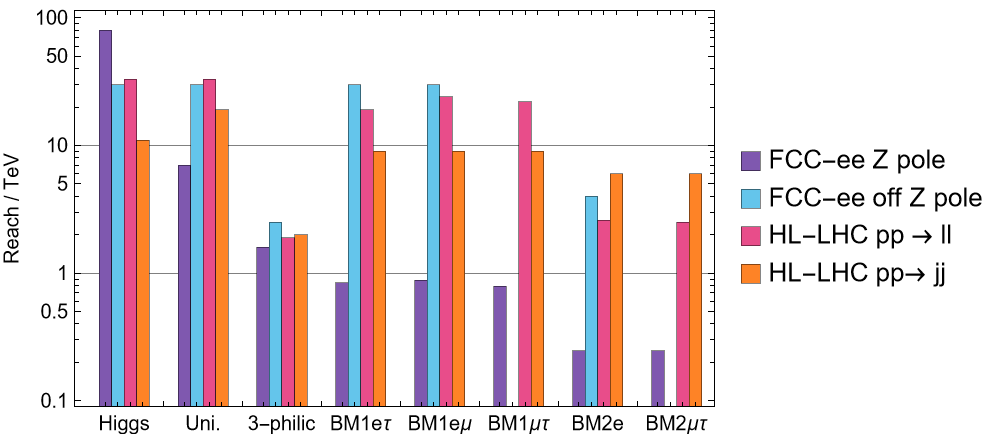}
    \caption{A comparison of the expected reach in $M/g$ (log-scale) of FCC-ee {\em vs.} HL-LHC for a set of benchmark $Z^\prime$ models which are designed to cover, qualitatively speaking, all anomaly-free $U(1)_X$-extensions of the SM. The first three scenarios are well-probed by EWPOs measured on the $Z$-pole at FCC-ee, as shown in~\cite{Allwicher:2024sso}. The other four scenarios correspond to $Z^\prime$ models that do not run into the EWPOs at 1-loop (leading log), as classified in this paper -- the weaker bounds on these models from tera-$Z$ come from small finite 1-loop matching contributions that cannot also be cancelled (\S \ref{sec:Zpole}). 
    All such models necessarily have order-1 coupling to valence quarks, and so strong LHC bounds cannot be avoided. BM1 has equal couplings to RH quarks and RH leptons; if electrons are charged the model will we probed up to around 30 TeV by FCC-ee runs above the $Z$-pole, while even if electrons are not charged, $pp\to\ell\ell$ measurements at the LHC already exclude all BM1 variants up to 10 TeV, with HL-LHC projections reaching above 20 TeV. 
    The final option BM2 has numerically tiny lepton charges, bringing down the constraints from Drell--Yan and FCC-ee. But, as we show, nothing that escapes the $Z$-pole at 1-loop can also escape the bounds from dijet searches at the LHC. We here estimate the sensitivity of HL-LHC in dijet searches by performing a na\"ive luminosity rescaling of the non-resonant search performed by CMS in~\cite{CMS:2018ucw}.  }
    \label{fig:BAR}
\end{figure}

We now put together our findings by comparing the dominant bounds for the benchmark $Z^\prime$ models considered in the main text, and also comparing against three benchmark scenarios studied in~\cite{Allwicher:2024sso} that {\em do} run into the EWPOs, in Fig.~\ref{fig:BAR}. As we have argued, the conditions of electroweak neutrality plus anomaly cancellation require an order-1 coupling to up quarks, which unavoidably results in strong bounds from the LHC experiments. The main phenomenological differences are then driven by which leptons are also charged (notably if electrons are charged, the $Z^\prime$ will be extremely well-probed at FCC {\em off} the $Z$-peak) and by whether the lepton charge is comparable to the quark charge or is numerically very small. 

The expected sensitivities that we plot in Fig.~\ref{fig:BAR} are for FCC-ee projections both on and off the $Z$-pole, where for the former we use~\cite{Allwicher:2024sso} in addition to the new computations in \S \ref{sec:Zpole}, and for the latter we use results from~\cite{Greljo:2024ytg}. For Drell--Yan we plot High-Luminosity LHC projections, which we compute using \texttt{HighPT}~\cite{Allwicher:2022mcg} as described in \S \ref{sec:2q2l}, which is based on extrapolating current bounds assuming HL-LHC will accumulate an integrated luminosity of 3 $\text{ab}^{-1}$.
For the dijet constraints we perform a similar na\"ive rescaling of the bound to an integrated luminosity of 3 $\text{ab}^{-1}$, using in this case the non-resonant search as quoted in \eqref{eq:4qEFTbound} for the case of BM2, for which a very slow improvement of $\Lambda$ scaling with the eighth root of the luminosity ratio is expected (similar to the non-resonant $pp\to \ell\ell$ case). An extrapolation of the resonant search result is harder to estimate, because the improvement of the dijet reach depends crucially on the kinematic endpoints of the search. For the first three benchmarks described below, which correspond to the scenarios in~\cite{Allwicher:2024sso}, the coupling to quarks is vector-like (and flavour-universal in two cases) rather than right-handed as for BMs 1 and 2, and so we use the bounds for a different signal model in the non-resonant dijet analysis of CMS~\cite{CMS:2018ucw} to perform our HL-LHC extrapolation.

The benchmarks we include in Fig.~\ref{fig:BAR}, and the corresponding bounds, are:
    \paragraph{`Higgs':} the $Z^\prime$ couples at tree-level to the Higgs and to all SM fermion generations, resulting in tree-level corrections to EWPOs. For instance, this could arise from gauging a linear combination of $B-L$ and hypercharge~\cite{Altmannshofer:2019xda}. 
    This scenario will be probed up to 80 TeV or so by FCC-ee $Z$-pole run. 
    Similarly strong $Z$-pole bounds would apply to less-minimal $U(1)_X$ extensions as long as the Higgs is charged, such as those associated with gauging a component of flavour non-universal hypercharge, {\em e.g.} in~\cite{Allanach:2018lvl,Allanach:2019iiy,Davighi:2023evx,FernandezNavarro:2023rhv,FernandezNavarro:2024hnv}.
    \paragraph{`Universal':} this toy $Z^\prime$ model does not couple to the Higgs, but couples universally to all SM fermion generations. An anomaly-free realisation of this scenario would be to gauge $U(1)_{B-L}$. It was shown that 1-loop running into EWPOs means this scenario will be probed to 8 TeV by the FCC-ee $Z$-pole run alone~\cite{Allwicher:2024sso}. But in fact, we observe that for this charge assignment showcased in~\cite{Allwicher:2024sso} the running into EWPOs with $y_t$ in fact cancels exactly, due to $u_3=q_3$, so this 8 TeV bound comes primarily from the running with $g_2$. We therefore point out that, for a chiral $Z^\prime$ model coupled differently to $u_3$ and $q_3$, the reach of the FCC-ee tera-$Z$ run would likely far exceed 8 TeV thanks to strong running with $y_t$.
    We supplement this with the bound recently obtained in~\cite{Greljo:2024ytg} by projecting FCC-ee {\em off}-peak constraints; to plot the ballpark bound of 30 TeV in Fig.~\ref{fig:BAR}, we take the bounds on 4-lepton operators that are flavour-conserving and probe the highest scales, which are $O_{ll,ee}^{1122}$, and $O_{ll,ee}^{1133}$, coming from $e^+e^-\to \mu^+\mu^- (\tau^+\tau^-)$ cross-section measurements and related observables. We see that the off $Z$-peak measurements actually probe higher scales than due to RGE running into the EWPOs, but this is in large part due to the cancellation of the $y_t$ running for this benchmark model as already noted.
    Since this model also has order-1 couplings to quarks, we perform a projection for the expected HL-LHC sensitivity in Drell--Yan $pp\to \ell\ell$ using \texttt{HighPT}~\cite{Allwicher:2022mcg}, including all three lepton flavour channels, which would already surpass 30 TeV even before the FCC-ee begins running. Lastly, for dijets we extrapolate the non-resonant bound from~\cite{CMS:2018ucw} relevant for a vector-like 4-quark operator, to obtain a strong projected sensitivity of 11 TeV.
    \paragraph{`$3^{\text{rd}}$ Gen':} this toy $Z^\prime$ model does not couple to the Higgs or to light fermions, but couples to all third generation SM fermions. An anomaly-free realisation of this scenario would be to gauge $U(1)_{B_3-L_3}$. The expected reach from EWPOs at FCC-ee was here shown to be 1.6 TeV from the 1-loop RGE running in~\cite{Allwicher:2024sso}, which is again subject to the caveat that the running with $y_t$ accidentally cancels exactly due to assigning $u_3=q_3$ for the benchmark: indeed, studies at the level of single SMEFT operator bounds show that pure third generation 4-fermion operators would get bounds of order $10\div 15$ TeV from the FCC-ee $Z$-pole run -- see {\em e.g.}~\cite{Allwicher:2023shc,Greljo:2024ytg}.
    For this third-family aligned $Z^\prime$ there are no tree-level bounds from the off $Z$-peak data because we do not couple to electrons, but at 1-loop the third generation 4-fermion operators will run into operators involving electrons, and so are still bounded by off-$Z$-peak measurements. In~\cite[Table VII]{Greljo:2024ytg} it was shown that these bounds reach 2.5 TeV, for operators ${O}_{dd}^{33}$ and $O_{qq}^{(1)\, 33}$, which we include in Fig.~\ref{fig:BAR}. 
    We also compute the sensitivity to this benchmark $Z^\prime$ model from $pp \to \tau\tau$ Drell--Yan measurements at the LHC, again using \texttt{HighPT}~\cite{Allwicher:2022mcg}.
    For this benchmark it is harder to estimate the HL sensitivity of the LHC dijet searches, because there is no EFT scenario in the non-resonant search that is third-family specific. In this case, 
    we extract a very crude current bound of $M_X/g_X \gtrsim 1$ TeV or so from the resonant search via the $b\bar{b}$ production mode using~\cite{Bordone:2021cca}, in a similar manner to \S \ref{sec:4q}. For the other models, we saw the bounds from non-resonant and resonant searches were similar in the region where both apply. If we assume similar holds for this third-generation specific scenario, we can na\"ively extrapolate with the gain in luminosity to obtain a crude projection of 1.5 TeV or thereabouts. 
    \paragraph{BM1:} our benchmark model 1 as introduced in the main text, \eqref{eq:BMs}. 
    The weak bounds from the $Z$ pole come from the small finite 1-loop matching terms derived in \S \ref{sec:Zpole}, and reach $\approx$ 850 GeV.
    We include bounds for three variants of this benchmark corresponding to permutations of the lepton flavour assignment; the key difference is that when $e_1 \neq 0$ there is a bound reaching 30 TeV from $e^+e^-\to \ell^+\ell^-$ off the $Z$ peak. 
    Note that even for the option with $e_1=0$ (named `BM1$\mu\tau$'), there will still be some sensitivity from off $Z$-pole FCC-ee data, due to finite 1-loop matching contributions to 4-fermion operators, similar to those computed in \S \ref{sec:Zpole}, but we do not show these bounds here.
    The HL-LHC Drell--Yan estimates are also slightly different for the three variants (\S \ref{sec:2q2l}), but all are roughly 20 TeV. We estimate the HL-LHC dijet non-resonant bound directly from the $\eta_{RR}$ signal model used by CMS in~\cite{CMS:2018ucw}, which is here appropriate because $X_u=X_d=1$, so we are able to set a slightly higher projection (than for BM2) of 9 TeV.
    \paragraph{BM2:} the second benchmark introduced in \eqref{eq:BMs} in which the lepton charge is very small compared to the quark charge. 
    Here the constraints from tera-$Z$, due to the small finite 1-loop matching effects, reach only 250 GeV.
    The most important bound in this case will likely come from dijet measurements, where we rescale the non-resonant bound \eqref{eq:4qEFTbound} with the luminosity gain to the eighth root to reach 6 TeV.
    Even though the lepton charge is very small, we find the expected sensitivity of HL-LHC Drell--Yan searches reaches nearly 2 TeV, driven by the dimuon channel. We also include a variant of BM2 (labelled BM2e in Fig.~\ref{fig:BAR}) in which electrons are charged for the sake of comparison, showing (from~\cite{Greljo:2024ytg}) that even with a numerically very small charge $|e_1| = |d_1|/41$ the off $Z$-peak measurements can reach above 4 TeV in sensitivity, largely thanks to the expected performance of bottom and charm tagging. 
From Fig.~\ref{fig:BAR} it is clear that there is excellent complementarity between $e^+e^-$ and $hh$ colliders for probing $Z^\prime$ models: in essence, the former puts strong constraints on the $U(1)_X$ coupling to leptons and to the Higgs, both through precision measurements at the $Z$-pole but also $e^+e^-\to f\bar{f}$ measurements off the $Z$-pole, while the latter constrains also the $U(1)_X$ couplings to quarks. Moreover, the conditions from requiring anomaly cancellation serve to tie the various charges together in an intricate way, so that there is no way to put the charges just into, say, heavy generation right-handed leptons or just into heavy flavour right-handed quarks, which would be the hardest scenario to detect. 

\section{Conclusion} \label{sec:conc}

In this paper, we sought to `fill a gap' left by~\cite{Allwicher:2024sso}, which demonstrated that almost all single particle extensions of the SM will be probed at least to the TeV scale purely through measuring EWPOs to very high precision, thanks to the mixing of operators under RG running of the SMEFT. 
Essentially the only exception that could be viable at low effective scales, considering also flavour probes and direct LHC bounds on coloured states, was found to be a family of $Z^\prime$ extensions of the SM.

We here close this gap by showing that, by combining the requirements from not running into the EWPOs with the requirements of gauge anomaly cancellation, any $Z^\prime$ that can escape the $Z$-pole cannot also escape the LHC, especially moving into the High-Luminosity phase. This is primarily because all such $Z^\prime$s must have order-1 couplings to up quarks. Anomaly cancellation also implies that at least two flavours of leptons must be charged.\footnote{We showed that none of these results are modified by the inclusion of an arbitrary number of right-handed neutrino fields with which we might try to `soak up' gauge anomalies.}
We emphasize that options in which electrons are directly charged would be probed up to 30 TeV by FCC-ee data {\em off} the $Z$-peak, surpassing the scales probed by the $Z$-pole due to RGE running; this reinforces the findings of the recent study~\cite{Greljo:2024ytg} that measurements above the $Z$-pole should form an integral part of the electroweak program at FCC-ee. Requiring the electron is neutral, we are able to classify all remaining anomaly-free $Z^\prime$ models analytically in terms of a pair of integers, up to certain permissible family index permutations.

Continuing in pursuit of an invisible $Z^\prime$, the two charged leptons that couple to the $Z^\prime$ ought to be muons and taus. 
In this case, Drell--Yan $pp\to \mu\mu,\, \tau\tau$ measurements from ATLAS and CMS put very strong bounds up to 10 TeV, which are projected to reach $20\div 25$ TeV after HL-LHC. In a last ditch attempt to evade also these bounds from $pp \to \ell\ell$, we are led to consider unusual anomaly-free models in which the lepton charge is numerically very small compared to the quark charge (which, incidentally, we prove cannot embed in any semi-simple UV completion). Even in this esoteric corner of model space, there remain strong constraints from LHC di-jet measurements that cannot be avoided, because the coupling to up quarks must be order-1. A conservative estimate of the current bound sits at 5 TeV. A na\"ive rescaling of the non-resonant bound on 4-quark operators coming from angular dijet analyses to the HL-LHC target integrated luminosity would give a projected sensitivity in the $6\div 10$ TeV range.

Reversing the argument: any anomaly-free $Z^\prime$ model that is {\em not} already constrained above the $10$ TeV mark after the LHC, for instance because it does not couple to light generation quarks, will necessarily run with $y_t$ and/or the SM gauge couplings into the $Z$-pole observables at 1-loop and so will be probed at FCC-ee. If the Higgs is charged under $U(1)_X$, it is strikingly clear from~\cite{Allwicher:2024sso} that the sensitivity of FCC-ee will be unrivalled.
Thus, in addition to closing the gap left by~\cite{Allwicher:2024sso}, we can make a broader conclusion, which is that there ought to be a perfect complementarity between a high-precision $e^+e^-$ collider (making the most of measurements both on and off the $Z$-pole) and the HL-LHC for probing generic anomaly-free $U(1)_X$ extensions of the SM gauge group.

\section*{Acknowledgement}

I am grateful to Lukas Allwicher, Marzia Bordone, Guilherme Guedes, Matthew McCullough, Sophie Renner, and Peter Stangl for discussions, and to Tim Cohen, Ayres Freitas, and Admir Greljo for their helpful comments on the manuscript. 
I am especially grateful to Lukas Allwicher for sharing code used to compute the bounds on $4l$ operators from LEP-II data in \S \ref{sec:4l}.

\bibliographystyle{JHEP}
\bibliography{refs}
\end{document}